\documentclass[]{iopjournal}

\usepackage{amsthm}

\theoremstyle{definition}
\newtheorem{definition}{Definition}

\newcommand{\PosInt}{\mathbb{N}}

\usepackage{amsmath,amssymb,amsfonts}
\usepackage{bm}
\usepackage{mathrsfs}
\usepackage{dsfont}
\usepackage{graphicx}
\usepackage{hyperref} % load last
\hypersetup{colorlinks=true,linkcolor=blue,citecolor=blue,urlcolor=blue}
\usepackage{rotating}

\usepackage{tikz}
\usepackage{tikz}
\usetikzlibrary{
    arrows.meta,
    positioning,
    fit,
    calc,
    shadings,
    fadings
}

\pgfdeclareradialshading{bluestate}{\pgfpoint{0cm}{0cm}}{
    color(0cm)=(blue!90!black);
    color(0.20cm)=(blue!85);
    color(0.45cm)=(blue!55);
    color(0.70cm)=(blue!20);
    color(1.00cm)=(white)
}

\tikzfading[
    name=statefade,
    inner color=transparent!0,
    outer color=transparent!100
]

\newcommand{\be}{\begin{equation}}
\newcommand{\ee}{\end{equation}}
\newcommand{\bea}{\begin{eqnarray}}
\newcommand{\eea}{\end{eqnarray}}

\newcommand{\ket}[1]{\left|#1\right\rangle}
\newcommand{\bra}[1]{\left\langle#1\right|}

\newcommand{\rev}[1]{#1}

\begin{document}

\articletype{Topical Review} %	 e.g. Paper, Letter, Topical Review...

\title{Quantum One-Way Functions and Related Cryptographic Primitives}

\author{Georgios M. Nikolopoulos$^{1,2,*}$\orcid{0000-0002-3937-2771}}

\affil{$^1$Institute of Electronic Structure \& Laser (IESL), FORTH, GR-70013 Heraklion, Greece}

\affil{$^2$Center for Quantum Science \& Technologies (FORTH-QuTech), GR-70013 Heraklion, Greece}

\affil{$^*$Author to whom any correspondence should be addressed.}

\email{nikolg@iesl.forth.gr}


\date{\today}

\begin{abstract}
\noindent
Quantum cryptographic primitives beyond key distribution remain a less well understood area of research. In classical cryptography, one-way functions underpin nearly all standard cryptographic protocols, motivating the search for meaningful quantum analogues and for a clear understanding of the physical and computational mechanisms that could enforce one-wayness. In this article, we review quantum one-way functions and a range of closely related quantum-state primitives, including one-way state generators, pseudorandom quantum states, and efficiently indistinguishable  pairs of states. We discuss both computational and information-theoretic notions of quantum one-wayness, emphasizing the different adversarial models and security assumptions that underlie these constructions. We compare and contrast the various proposed primitives, and clarify their conceptual relationships. 
Particular emphasis is placed on questions of physical realizability, experimental feasibility, and robustness to noise. Finally, we outline open problems and future directions toward the development of practical quantum cryptographic primitives beyond key distribution, and the emergence of a broader quantum-cryptographic ecosystem.
\end{abstract}

\section{Introduction}
\label{sec:intro}

To date, quantum cryptography has been largely synonymous with quantum key distribution (QKD) 
\cite{QKD-RMP02,Scarani-RMP09,LoQKD_NP14,Diamanti2016PracticalQKD,Xu-RMP20,QC-RMP25,QKDNet-ACM25,NielsenChuang2000,Wolf2021QKD}, which remains the most mature and widely deployed application of quantum information science. 
QKD has demonstrated that, by exploiting fundamental principles of quantum physics, it is possible to achieve cryptographic tasks that are unattainable in purely classical settings. 
However, the broad proliferation of quantum cryptography and the long-term vision of a quantum internet motivate the development of additional cryptographic primitives and functionalities, 
such as digital signatures, authentication mechanisms, and encryption schemes, that go beyond key distribution.

In the classical setting, such primitives are built upon mathematical one-way functions (OWFs), which are widely regarded as the cornerstone of modern cryptography \cite{GoldreichFoC1,GoldreichFoC2}. 
The assumed existence of OWFs enables the construction of nearly all standard cryptographic primitives. 
This naturally raises the question of whether analogous notions of one-wayness exist in the quantum domain. 
What physical principles can prevent the inversion of quantum one-way functions (QOWFs)
\footnote{The term QOWF is used inconsistently in the literature. In parts of the theoretical-cryptography community it often refers to a  post-quantum one-way function with classical output, whereas in the quantum-information literature it usually denotes a mapping involving quantum states. Throughout this review we adopt the latter convention. This difference in terminology is one of the motivations for the present review.}?
What are the requirements for their realization, and what are the fundamental limits of their applicability?

A growing body of literature has addressed aspects of these questions, yet several conceptual and practical issues remain unresolved. 
Various quantum-state primitives associated with one-wayness, indistinguishability, and pseudorandomness, such as QOWFs \cite{gottesman2001}, one-way state generators (OWSGs) \cite{MorimaeYamakawaTQC2024}, pseudorandom quantum state generators (PRSGs) \cite{JiLiuSong2018}, and efficiently indistinguishable (EFI) pairs \cite{KKNY2013}, have been proposed as candidate foundational building blocks for quantum cryptography. 
At the same time, these constructions exhibit important differences from mathematical OWFs, particularly with respect to their security assumptions and physical realizability. 
In contrast to purely mathematical OWFs, quantum constructions often rely on physical constraints, limited access models, or bounded numbers of quantum states, which introduce new challenges and limitations.

The primitives discussed in this review originated independently in partially overlapping research communities, and were often introduced using different terminology, motivations, and security frameworks. 
Early work in quantum information focused on classical-to-quantum one-way functions and physically motivated notions of one-wayness, whereas more recent developments in theoretical cryptography introduced related concepts such as OWSGs, EFI state pairs, and PRSGs. 
Although these primitives differ in their formal definitions and intended applications, they rely on  the difficulty of extracting, reproducing, distinguishing, or otherwise characterizing quantum information. 
The purpose of this review is not to provide a comprehensive survey of all quantum cryptographic assumptions, protocols, and applications, but rather to organize and compare those primitives 
that are directly connected to quantum one-wayness and quantum-state generation, while highlighting conceptual links between developments that have emerged independently in the quantum-information and theoretical-cryptography communities.

The focus of this review is on quantum cryptographic primitives whose operation or security fundamentally relies on quantum information processing, such as the preparation, transmission, or measurement of quantum states. 
This should be distinguished from post-quantum cryptography, which studies entirely classical schemes that remain secure against quantum adversaries under computational hardness assumptions. 
Our aim is threefold: (i) to provide a structured overview of the main results concerning quantum one-way functions and related constructions; (ii) to discuss their relationship as well as their differences; and (iii) to highlight open problems and outline a roadmap for future research in this area. We emphasize that the present review is not intended to provide a comprehensive survey of all candidate cryptographic assumptions and primitives appearing in the broader Microcrypt landscape, together with their reductions, separations, and applications. For such a perspective, the reader may consult resources such as the Microcrypt Zoo~\cite{MicrocryptZoo}. Instead, we focus on QOWFs and a small number of closely related primitives whose security or functionality fundamentally relies on quantum states. 
While post-quantum assumptions may sometimes be used as building blocks to obtain computational security for quantum primitives, the focus of this review is on constructions in which the cryptographic objects themselves are quantum. The article is written primarily for readers with a background in physics who are interested in quantum cryptography, but who may not be familiar with the standard computational-complexity formalism. To make the presentation accessible to a broad readership, we deliberately keep technical formalism to a minimum while emphasizing physical intuition and conceptual clarity.

The primitives reviewed in this article emphasize different aspects of quantum one-wayness and quantum-state generation. QOWFs and OWSGs focus on the difficulty of recovering classical information encoded in quantum states. EFI pairs capture computational indistinguishability between efficiently preparable quantum states, while PRSGs strengthen this notion by requiring computational pseudorandomness. Although these primitives differ in their formal definitions, security assumptions, and intended applications, they can be viewed as complementary approaches to exploiting uniquely quantum properties for cryptographic purposes. Their relationships and distinguishing features are summarized in Table~\ref{tab:primitives}. Together, they provide building blocks for a broader framework of quantum cryptography and give rise to a variety of cryptographic applications, including quantum money, authentication schemes, quantum public-key constructions, and quantum encryption protocols. A detailed discussion of such applications lies beyond the scope of the present review.

The remainder of this article is organized as follows. 
In section \ref{sec:owf}, we briefly review classical OWFs and their role in modern cryptography, establishing the conceptual framework that motivates their quantum counterparts. 
Section \ref{sec:qowf} introduces QOWFs and OWSGs, focusing on classical-to-quantum maps. 
In section \ref{sec:ind}, we discuss possible connections of quantum one-wayness to indistinguishability, while EFI pairs and PRSGs are discussed in 
sections \ref{sec:efi} and \ref{sec:prsg}, respectively. 
Section \ref{sec:other} is devoted to various constructions of QOWFs that go beyond classical-to-quantum maps, while in section \ref{sec:puf}, we examine physically motivated notions of one-wayness, with a particular focus on classical and quantum physical unclonable functions (PUFs), highlighting their similarities to and fundamental differences from QOWFs. Section \ref{sec:qowf_pqc} discusses the role of quantum one-wayness in the era of post-quantum cryptography. 
Finally, in section \ref{sec:outlook}, we summarize the main insights of the review and outline open problems and future directions toward a broader theory of quantum one-wayness and related quantum cryptographic primitives.

\section{One-way functions}
\label{sec:owf}

One-way Functions (OWFs) are efficiently computable functions that are hard to invert on average \cite{GoldreichFoC1}. Formally, a function
\[
f:\{0,1\}^* \mapsto \{0,1\}^*
\]
is called a one-way function if (i) there exists a polynomial-time algorithm that computes $f(x)$ for any input $x$, but (ii) for a given random $y=f(x)$,  the probability for every probabilistic polynomial-time (PPT) adversary ${\cal A}$ \footnote{A PPT adversary is a realistic attacker modeled as a randomized algorithm (i.e., it can make random choices) 
  and is limited to running in a number of computational steps that grows at most as a polynomial in the security parameter  (equivalently, polynomial in the input size when the input encodes the security parameter).}
to return a valid preimage $x'$ such that $f(x') = y$, is negligible in the input length. 
It is important to note here that in the second condition the preimage that is returned by the algorithm need not be the actual $x$. If $f$ is injective, then recovering any preimage is essentially the same as recovering the unique $x$ such that  $f(x) = y$. Moreover, the hardness of finding a preimage,  
does not mean that it is infeasible to find some partial information about the preimage of $y$ under $f$ (e.g., see related discussion in section 2.5 of \cite{GoldreichFoC1}).

OWFs are widely regarded as the minimal computational assumption underlying much of modern cryptography. This is because the existence of pseudorandom generators (PRGs) is equivalent to the existence of OWFs \cite{HILL1999,GoldreichFoC2}. In turn, pseudorandomness is a central ingredient in most computationally secure cryptographic schemes. Consequently, in the absence of OWFs, only a limited number of information-theoretic primitives, such as the one-time pad or secret sharing, remain feasible. At the same time, OWFs are remarkably powerful: their existence suffices to construct a wide range of cryptographic primitives, including PRGs, pseudorandom functions, symmetric-key encryption schemes, message authentication codes, commitment schemes, and digital signatures. 

OWFs do not immediately yield public-key encryption, which is not known to be constructible from OWFs alone and typically requires stronger assumptions, such as trapdoor OWFs. A trapdoor OWF is a special type of OWF that is easy to compute in the forward direction, but hard to invert unless one possesses some secret piece of information called the \emph{trapdoor}. Given the trapdoor, inversion becomes efficient, whereas without it the function remains one-way. Trapdoor OWFs enable public-key primitives such as encryption. At the same time, it is worth noting that digital signatures can in principle be constructed from OWFs alone, for example through Lamport’s one-time signature scheme, although practical signature schemes often rely on stronger assumptions. Since every trapdoor OWF is also an OWF, the latter remain the fundamental minimal building block underlying much of modern cryptography \cite{GoldreichFoC2,KatzLindell}.

While OWFs are abstract complexity-theoretic notions, their concrete instantiations in cryptography almost always rely on mathematical hardness assumptions, such as factoring, discrete logarithms, lattice problems, or coding-theoretic problems. The hardness of these problems, however, remains unproved. In fact, Shor showed that some widely used assumptions, including factoring and discrete logarithms, can be solved in polynomial time on a quantum computer \cite{Shor1997}. This observation motivates the search for post-quantum alternatives.

\subsection{Post-quantum one-way functions}

A  post-quantum OWF  is an ordinary classical function 
$f:\{0,1\}^* \mapsto \{0,1\}^*$ that is efficiently computable, but conjectured to be 
hard to invert for any \emph{quantum} polynomial-time adversary (QPT)\footnote{A QPT adversary is a quantum algorithm whose running time is polynomially bounded in the security parameter, and which may make use of arbitrary quantum superposition states, entangled states, and measurements during its execution.}. In other words, 
the definition of one-wayness is the same as in the classical setting, except that 
the adversary is allowed to be a quantum algorithm. 

These functions form the foundation of post-quantum cryptography (PQC). 
Many of the fundamental relationships known from classical cryptography continue to hold in PQC. 
In particular, post-quantum OWFs can be used to construct PRGs and other cryptographic primitives 
that remain secure against quantum adversaries, although the corresponding reductions often 
require additional technical care to account for quantum queries.

Unlike some traditional OWF candidates based on factoring or discrete logarithms, which are broken by 
Shor's quantum algorithm~\cite{Shor1997}, current candidate post-quantum OWFs rely on the 
hardness of problems that are believed to withstand quantum attacks, such as 
\emph{Learning With Errors} (LWE) and other lattice-based problems, 
code-based problems, and assumptions underlying hash-based constructions~\cite{GoldreichFoC2,KatzLindell}. 
Thus, post-quantum OWFs are generally regarded as one of the minimal computational assumptions underlying much of PQC.

\subsection{One-wayness and indistinguishability} 

As mentioned above, OWFs are defined in terms of a \emph{search problem} i.e., given an output $y=f(x)$, it should be computationally infeasible 
to recover a valid input $x'$ such that $y=f(x')$. By contrast, many cryptographic primitives are naturally phrased as \emph{decision problems}, where 
the task is to distinguish between two possible distributions. 

The notion of \emph{computational indistinguishability} plays a pivotal role in modern cryptography. 
Rigorous definitions and extensive discussions can be found in standard textbooks \cite{GoldreichFoC1,KatzLindell}.
For the purposes of the present review, it is sufficient to keep in mind that two probability
distributions are computationally indistinguishable if no efficient (PPT) algorithm (the so-called  distinguisher)
can tell them apart. In other words, the distinguisher ${\cal D}$ is  given a sample from one of the distributions. The probability for the 
distinguisher in deducing correctly the distribution from which the sample has been obtained, is  only negligibly better than random guessing.

Computational indistinguishability is weaker than statistical closeness (sometimes also referred to as variation distance) between the two pertinent distributions. 
If two distributions are statistically extremely close (i.e., their statistical distance is negligible), 
then  no polynomial-time observer can distinguish them. The converse implication, however, does not generally hold. In particular, there exist distributions that are not statistically close and yet remain computationally indistinguishable \cite{GoldreichFoC1,GoldreichFoC2,KatzLindell}. 

A remarkable feature of modern cryptography is that, under suitable conditions, search hardness can be transformed into decision hardness. In particular, the existence of OWFs implies the existence of hard-core predicates 
and pseudorandom generators \cite{GoldreichLevin1989,HILL1999}. Informally, a hard-core predicate is a single bit of information that remains computationally hidden even when the value of the OWF is known. 
Such constructions establish a connection between one-wayness and computational indistinguishability, and play a central role in the construction of pseudorandom objects used throughout modern cryptography.

In closing this brief summary of mathematical OWFs, it is worth keeping in mind that 
OWFs can be used as foundational building blocks for a wide range of
cryptographic tasks, and the typical chain of operations is as follows.
Classical information is processed using cryptographic primitives that
are ultimately constructed from OWFs, producing classical outputs which,
together with any additional information required by the protocol, are
sent to the receiver over a classical channel. The receiver performs the
corresponding verification or decryption procedure to complete the task.
Quantum cryptography does not eliminate the classical world. Instead, it
introduces quantum steps that provide new security guarantees, such as
unclonability and measurement disturbance. Nevertheless, in many cryptographic applications the information 
ultimately being communicated, authenticated, or verified remains classical, even when quantum states themselves become part of the cryptographic protocol.

\section{Quantum one-way functions}
\label{sec:qowf} 

This section introduces the main notions of quantum one-wayness and reviews representative QOWF constructions.

\subsection{Physical principles and constraints governing QOWFs}

Unlike classical OWFs, QOWFs involve quantum states of physical systems and are therefore constrained by the laws of quantum physics. As a result, quantum one-wayness may arise not only from computational hardness assumptions, but also from fundamental physical limitations on the manipulation, measurement, and transmission of quantum information. Depending on the construction, security may rely on either or both of these ingredients. We briefly review some of the key physical principles that motivate the study of quantum one-wayness.

{\em 1. Quantum measurement.} In quantum physics, perfect discrimination between non-orthogonal states is impossible \cite{NielsenChuang2000,Barnett2009}.
If the output of a QOWF is a quantum state that belongs to a set of non-orthogonal states, then 
even an adversary with unlimited computational resources cannot perfectly learn the output state, and any measurement strategy incurs a nonzero
probability of error. 
The outcomes of  quantum measurements are fundamentally probabilistic and inherently random 
(unless the system is in an eigenstate of the operator being measured).
%This property underlies the information-theoretic security of many QOWFs, such as Wiesner's quantum money scheme, where banknote qubits are prepared in random non-orthogonal bases~\cite{wiesner1983conjugate}. 
The design of measurements that minimize error probability or maximize information gain has been studied extensively in the 
literature for many years \cite{Barnett2009,BarnettCroke2009,NikAJP25}. 
Optimal measurements are known for certain scenarios only (e.g., distinguishing between symmetric pure states). 

{\em 2. Information gain implies disturbance.}
Extracting information from a quantum system requires measurement, which  
will in general disturb the state of the system in a way that limits
subsequent information extraction \cite{NielsenChuang2000,Barnett2009}. 
This implies that an adversary cannot in general perform multiple independent measurements  on the same unknown state, 
in contrast to classical settings where the output of an OWF can be copied and analyzed repeatedly. 
As a result, inversion of a QOWF is inherently constrained by the destructive nature of quantum measurements.

{\em 3. Holevo's theorem.}
The amount of classical information that can be extracted from a single $d-$dimensional quantum system is bounded from above by the logarithm of its Hilbert space dimension \cite{holevo1973bounds,NielsenChuang2000},
\begin{equation}
    I_{\text{acc}} \leq \log_2 (d).
\end{equation}
For example, a single qubit ($d=2$) carries at most one bit of extractable classical information, regardless of how many classical bits were encoded in its state. 
Consequently, if a QOWF compresses a large classical input into a lower-dimensional quantum state, the inversion problem is information-theoretically constrained, in the
sense that not all encoded classical information can be reliably
recovered.

{\em 4. No-cloning theorem.}
In classical cryptography, one can create multiple copies of the output of an OWF and attempt multiple inversion strategies. 
In quantum mechanics, however, this is forbidden by the no-cloning theorem~\cite{wootters1982nocloning}.   
The impossibility of creating perfect copies of an unknown quantum state
can strengthen one-wayness in certain QOWF constructions, since even
probabilistic inversion strategies are fundamentally restricted.

These physical principles jointly provide the foundations for the design of QOWFs, but at the same time they impose several  challenges with respect to their realization  in practice. 
Quantum states are very fragile, and they are susceptible to decoherence, imperfections and environmental noise. Despite the tremendous progress that has been achieved in the context of QKD and other quantum technologies,  
the generation and manipulation of complex quantum states remains challenging, while the reliable long-term storage of quantum states remains an open issue. 

In the classical setting, many operations are taken for granted. For instance, one can readily compare the outputs of an OWF for two distinct inputs in order to determine whether $f(x)=f(x')$. The comparison of quantum states, however, generally requires quantum operations and measurements, and may not be implementable with perfect reliability. Unless the states are orthogonal, any comparison procedure is subject to a nonzero probability of error as a consequence of the fundamental principles discussed above \cite{gottesman2001,Buhrman2001,NikolopoulosIoannouPRA2009}. Such considerations illustrate that the design of QOWFs is highly nontrivial, particularly when moving beyond conceptual security to questions of physical implementability.

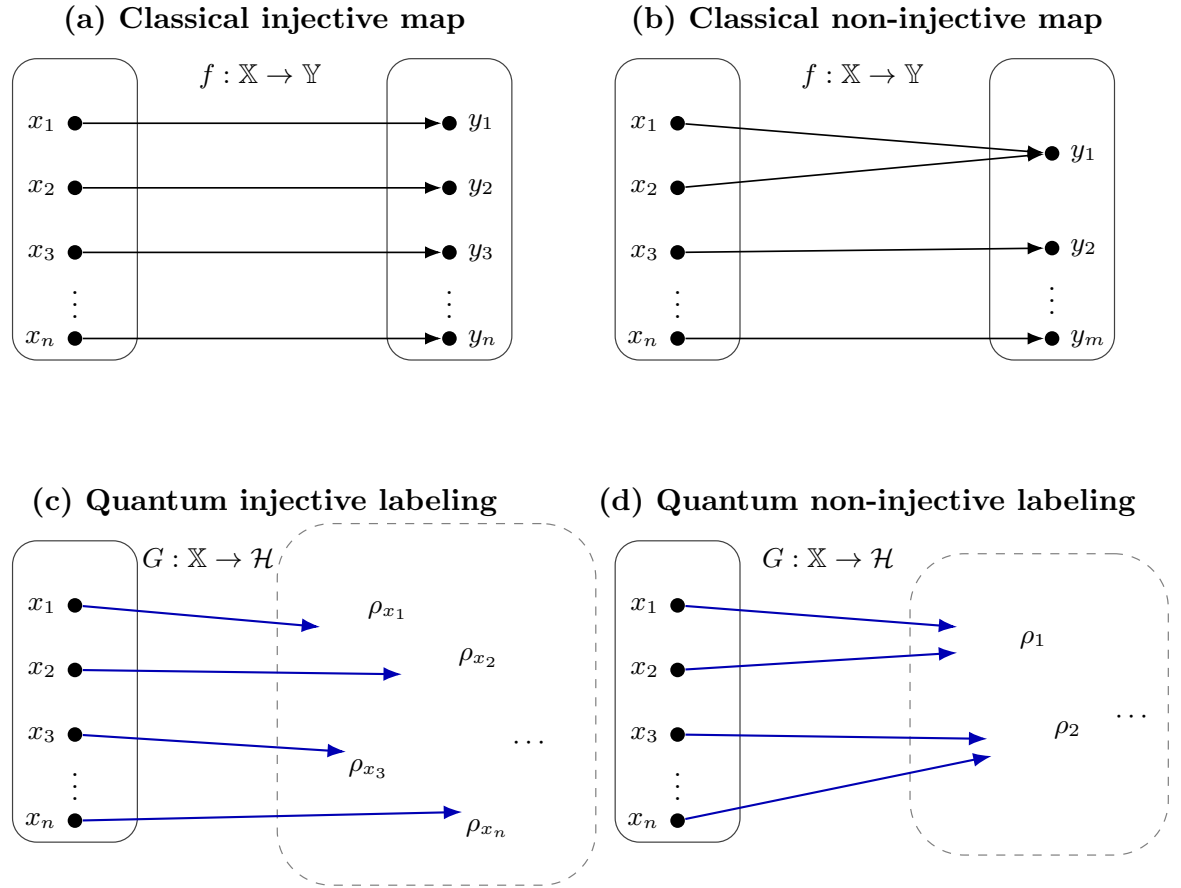
\begin{figure}[t]
\centering

\resizebox{\textwidth}{!}{%
\begin{tikzpicture}[
    font=\small,
    >=Latex,
    maparrow/.style={
        ->,
        line width=0.55pt
    },
    qarrow/.style={
        ->,
        line width=0.7pt,
        draw=blue!70!black
    },
    point/.style={
        circle,
        fill=black,
        inner sep=1.7pt
    },
    setbox/.style={
        draw=black!75,
        rounded corners=8pt,
        minimum width=1.45cm,
        minimum height=3.5cm
    },
    statespace/.style={
        draw=black!50,
        dashed,
        rounded corners=18pt,
        minimum width=3.7cm,
        minimum height=4.2cm
    },
      statespacev2/.style={
        draw=black!50,
        dashed,
        rounded corners=18pt,
        minimum width=3.cm,
        minimum height=3.5cm
    },
    state/.style={
        circle,
        minimum size=1.75cm,
        inner sep=0pt,
        shading=bluestate,
        path fading=statefade,
        draw=none
    },
    panel/.style={
        draw=black!55,
        rounded corners=3pt,
        inner sep=7pt
    }
]

% ============================================================
% (a) Classical injective map
% ============================================================

\begin{scope}[local bounding box=panelA]

\node[font=\bfseries\normalsize] at (3.05,4.15)
    {(a) Classical injective map};

\node[setbox] (XA) at (0.85,1.95) {};
\node[setbox] (YA) at (5.20,1.95) {};

%\node[above=3pt of XA] {$X$};
%\node[above=3pt of YA] {$Y$};

\node at (3.00,3.50) {$f:{\mathbb X}\rightarrow {\mathbb Y}$};

\node[point,label=left:$x_1$] (xa1) at (0.85,2.95) {};
\node[point,label=left:$x_2$] (xa2) at (0.85,2.20) {};
\node[point,label=left:$x_3$] (xa3) at (0.85,1.45) {};
\node at (0.85,0.95) {$\vdots$};
\node[point,label=left:$x_n$] (xan) at (0.85,0.45) {};

\node[point,label=right:$y_1$] (ya1) at (5.20,2.95) {};
\node[point,label=right:$y_2$] (ya2) at (5.20,2.20) {};
\node[point,label=right:$y_3$] (ya3) at (5.20,1.45) {};
\node at (5.20,0.95) {$\vdots$};
\node[point,label=right:$y_n$] (yan) at (5.20,0.45) {};

\draw[maparrow] (xa1) -- (ya1);
\draw[maparrow] (xa2) -- (ya2);
\draw[maparrow] (xa3) -- (ya3);
\draw[maparrow] (xan) -- (yan);

\end{scope}

%\node[panel,fit=(panelA)] {};

% ============================================================
% (b) Classical non-injective map
% ============================================================

\begin{scope}[xshift=7.0cm,local bounding box=panelB]

\node[font=\bfseries\normalsize] at (3.05,4.15)
    {(b) Classical non-injective map};

\node[setbox] (XB) at (0.85,1.95) {};
\node[setbox] (YB) at (5.20,1.95) {};

%\node[above=3pt of XB] {${\mathbb X}$};
%\node[above=3pt of YB] {${\mathbb Y}$};

\node at (3.00,3.50) {$f:{\mathbb X}\rightarrow {\mathbb Y}$};

\node[point,label=left:$x_1$] (xb1) at (0.85,2.95) {};
\node[point,label=left:$x_2$] (xb2) at (0.85,2.20) {};
\node[point,label=left:$x_3$] (xb3) at (0.85,1.45) {};
\node at (0.85,0.95) {$\vdots$};
\node[point,label=left:$x_n$] (xbn) at (0.85,0.45) {};

\node[point,label=right:$y_1$] (yb1) at (5.20,2.60) {};
\node[point,label=right:$y_2$] (yb2) at (5.20,1.50) {};
\node at (5.20,1.00) {$\vdots$};
\node[point,label=right:$y_m$] (ybm) at (5.20,0.45) {};

\draw[maparrow] (xb1) -- (yb1);
\draw[maparrow] (xb2) -- (yb1);
\draw[maparrow] (xb3) -- (yb2);
\draw[maparrow] (xbn) -- (ybm);

\end{scope}

%\node[panel,fit=(panelB)] {};

% ============================================================
% (c) Quantum injective labeling
% ============================================================

\begin{scope}[yshift=-5.6cm,local bounding box=panelC]

\node[
    font=\bfseries\normalsize,
    text=black
] at (3.05,4.15)
    {(c) Quantum injective labeling};

\node[setbox] (XC) at (0.85,1.95) {};
\node[statespace] (HC) at (5.05,1.8) {};

%\node[above=3pt of XC] {$X$};
%\node[above=3pt of HC] {State space $\mathcal{H}$};

\node at (2.4,3.50) {$G:{\mathbb X}\rightarrow\mathcal{H}$};

\node[point,label=left:$x_1$] (xc1) at (0.85,2.95) {};
\node[point,label=left:$x_2$] (xc2) at (0.85,2.20) {};
\node[point,label=left:$x_3$] (xc3) at (0.85,1.45) {};
\node at (0.85,0.95) {$\vdots$};
\node[point,label=left:$x_n$] (xcn) at (0.85,0.45) {};

% Quantum-state peaks
\node[state] (qc1) at (4.20,2.70) {};
\node[state] (qc2) at (5.25,2.15) {};
\node[state] (qc3) at (4.55,1.25) {};
\node[state] (qcn) at (5.95,0.55) {};

% Labels are separate so that they are not affected by fading
\node[above right=-1pt and -2pt of qc1.center] {$\rho_{x_1}$};
\node[above right=-1pt and -2pt of qc2.center] {$\rho_{x_2}$};
\node[below left=-1pt and -2pt of qc3.center] {$\rho_{x_3}$};
\node[below left=-1pt and -2pt of qcn.center] {$\rho_{x_n}$};

\node at (6.15,1.35) {$\cdots$};

\draw[qarrow] (xc1) -- (3.70,2.70);
\draw[qarrow] (xc2) -- (4.65,2.15);
\draw[qarrow] (xc3) -- (4.00,1.25);
\draw[qarrow] (xcn) -- (5.35,0.55);

\end{scope}

%\node[panel,fit=(panelC)] {};

% ============================================================
% (d) Quantum non-injective labeling
% ============================================================

\begin{scope}[
    xshift=7.0cm,
    yshift=-5.6cm,
    local bounding box=panelD
]

\node[
    font=\bfseries\normalsize,
    text=black
] at (3.05,4.15)
    {(d) Quantum non-injective labeling};

\node[setbox] (XD) at (0.85,1.95) {};
\node[statespacev2] (HD) at (5.05,1.8) {};

%\node[above=3pt of XD] {$X$};
%\node[above=3pt of HD] {State space $\mathcal{H}$};

\node at (2.6,3.50) {$G:{\mathbb X}\rightarrow\mathcal{H}$};

\node[point,label=left:$x_1$] (xd1) at (0.85,2.95) {};
\node[point,label=left:$x_2$] (xd2) at (0.85,2.20) {};
\node[point,label=left:$x_3$] (xd3) at (0.85,1.45) {};
\node at (0.85,0.95) {$\vdots$};
\node[point,label=left:$x_n$] (xdn) at (0.85,0.45) {};

\node[state] (qd1) at (4.70,2.55) {};
\node[state] (qd2) at (5.10,1.5) {};

\node[right=0pt of qd1.center] {$\rho_1$};
\node[right=0pt of qd2.center] {$\rho_2$};

\node at (6.15,1.65) {$\cdots$};

\draw[qarrow] (xd1) -- (4.10,2.70);
\draw[qarrow] (xd2) -- (4.10,2.40);
\draw[qarrow] (xd3) -- (4.45,1.40);
\draw[qarrow] (xdn) -- (4.50,1.20);

\end{scope}

%\node[panel,fit=(panelD)] {};

\end{tikzpicture}%
}

\caption{Schematic comparison of injective and non-injective mappings in
the classical and quantum settings. In the quantum case, the shaded regions
represent states with finite operational overlap; distinct labels may therefore
correspond to distinct, but nonorthogonal states.}
\label{fig:classical-quantum-mappings}
\end{figure}

\subsection{Classical-to-quantum one-way functions}
\label{sec:qowf_part2}

One natural approach to quantum one-wayness is to encode classical information into quantum states. In this setting, a classical input is used to generate a quantum state that serves as the cryptographic object. Such constructions are particularly attractive because they combine classical descriptions and verification procedures with uniquely quantum features, such as limited distinguishability, measurement disturbance, and unclonability. They naturally follow a classical–quantum–classical workflow: classical information specifies the state preparation, quantum information provides the cryptographic resource, and classical information is ultimately recovered or verified. As in classical cryptography, these constructions are defined with respect to a security parameter. For each value of the security parameter, the corresponding family of quantum states and preparation procedures is finite and efficiently describable, allowing both security analysis and physical implementation to be formulated in a well-defined manner. Although an adversary may in principle prepare or manipulate arbitrary states within the full Hilbert space associated with the underlying quantum system, the cryptographic construction itself is specified by a finite family of efficiently describable states indexed by the security parameter.

Many QOWF constructions considered in the literature are of the form ${\mathbb X} \mapsto \mathbb{S}$, where an input domain  $ {\mathbb X} $ (consisting of integers, binary strings, etc) is mapped onto a set ${\mathbb S}$ of distinct quantum (in general mixed) states. The set ${\mathbb S}$ is a subset of the Hilbert space  associated with a $d-$dimensional  quantum system, and $n:=\lceil \log_2|\mathbb{X}|\rceil $ is the number of bits required for indexing the input domain. 

The notion of quantum one-wayness encompasses several distinct mechanisms that may prevent an adversary from successfully inverting a classical-to-quantum mapping. Throughout the literature, the term ``one-way'' is often used in contexts that differ significantly with respect to the inversion objective, the adversarial model, the available resources, and the underlying security assumptions. Since the present review discusses constructions originating from both the cryptographic and quantum-information communities, it is useful to distinguish these aspects explicitly.

We formulate quantum one-wayness in the framework of an inversion game between two entities, the challenger and the adversary. The challenger samples a uniformly random $x\in{\mathbb X}$ and prepares $c\geq1$ independent copies of the corresponding quantum state $\rho_x$. The adversary receives the joint state $\rho_x^{\otimes c}$ together with the security parameter $n$. The parameter $c$ specifies the number of copies of the challenge state available to the adversary and plays a central role in many notions of quantum one-wayness.
We denote by $\mathsf{Succ}(x,x')$ the event that the adversary succeeds when the
challenge label is $x$ and the adversary outputs $x'\in\mathbb X$. Two natural recovery objectives arise in this setting. 
\begin{enumerate}
\item[(i)] \textbf{Exact-label recovery.}
The adversary succeeds only if it outputs the original classical label,
\[
\mathsf{Succ}(x,x')
\iff
x'=x.
\]

\item[(ii)] \textbf{Approximate inversion.}
The adversary succeeds if it outputs a label $x'\in{\mathbb X}$ such that the corresponding state $\rho_{x'}$ is sufficiently close to the challenge state $\rho_x$ according to a prescribed similarity measure. Throughout this review, approximate inversion is formulated using the Uhlmann fidelity, and the adversary succeeds whenever
\[
\mathsf{Succ}(x,x')
\iff
F(\rho_x,\rho_{x'})\ge\tau,
\]
where $\tau\in(1/2,1)$ is a fixed known constant. Other standard measures of similarity or distance between quantum states, such as the trace distance, could also be employed.
\end{enumerate}

\rev{
Exact-label recovery asks whether the particular classical label sampled by
the challenger can be identified. Approximate inversion, by contrast, asks
whether the adversary can recover any label whose associated quantum state
satisfies the prescribed operational acceptance criterion. As discussed
below, the former objective has been used in several constructions described
in the literature as quantum one-way functions, but it is not by itself a
direct analogue of classical preimage resistance, especially for
non-injective mappings.
}

Recall that for two quantum states $\rho$ and $\sigma$, the (Uhlmann)
fidelity is defined as \cite{NielsenChuang2000,Barnett2009}
\begin{equation}
F(\rho,\sigma)=
\left(
\mathrm{Tr}
\sqrt{\sqrt{\rho}\sigma\sqrt{\rho}}
\right)^2,
\end{equation}
and provides a standard measure of similarity between quantum states,
taking values between $0$ and $1$, where $F=1$ if and only if the states are identical.
For pure states, $\rho=\ket{\psi}\!\bra{\psi}$ and
$\sigma=\ket{\phi}\!\bra{\phi}$, this expression reduces to
\[
F(\psi,\phi)=|\langle\psi|\phi\rangle|^2.
\]

\rev{
The distinction between exact-label recovery and operational inversion is
particularly important for classical-to-quantum mappings, as illustrated
schematically in Fig.~\ref{fig:classical-quantum-mappings}.
For a classical function $f:{\mathbb X}\rightarrow {\mathbb Y}$, inversion of an output
$y=f(x)$ means finding any valid preimage $x'$ satisfying $f(x')=y$.
If $f$ is injective, the valid preimage is unique and therefore exact-label
recovery and inversion coincide. For a non-injective classical function,
however, several distinct inputs may constitute valid preimages.
}

\rev{
The quantum setting introduces an additional distinction. Even when the
classical labeling $x\mapsto\rho_x$ is injective, so that
$x\neq x'$ implies $\rho_x\neq\rho_{x'}$, distinct output states need not
be orthogonal and may have substantial overlap. Hence an incorrect label
$x'\neq x$ may nevertheless generate a state $\rho_{x'}$ that is accepted
by an operational verification procedure. Injectivity of the classical
labeling therefore does not imply the existence of a unique operationally
acceptable inverse. This motivates distinguishing exact recovery of the
preparation label from approximate inversion, in which any label satisfying
the prescribed verification criterion is regarded as successful.
}

\rev{
This distinction is also important for interpreting exact-label hardness.
For a non-injective mapping, difficulty in identifying the particular label
sampled by the challenger does not necessarily imply preimage resistance.
For example, a constant mapping $x\mapsto\rho$ may make recovery of a
uniformly sampled $n$-bit label possible only with probability $2^{-n}$,
while inversion is trivial because every label produces the same output.
Exact-label recovery should therefore be regarded as a separate
label-identification problem rather than, by itself, as a general analogue
of classical one-wayness.
}

\rev{
From a physical perspective, many constructions historically described as
quantum one-way functions were introduced as components of specific
cryptographic protocols, and their operational meaning is therefore tied
to the verification procedure of the application. Nonzero overlap between
states associated with distinct labels is often unavoidable and does not,
by itself, constitute a successful attack. Protocols may employ repeated
state preparations or measurements and base their final decision on the
resulting statistics, thereby amplifying the separation between legitimate
and incorrect responses. This helps explain why exact recovery of the
classical preparation label has frequently been studied as a notion of
quantum one-wayness in the physics literature. At the level of the primitive,
however, exact-label recovery and operational preimage resistance should be
kept conceptually distinct.
}

\rev{
Combining either recovery objective with either a computational
or an information-theoretic adversarial model gives rise to four corresponding
security notions. These notions need not all constitute one-wayness in the classical
preimage-resistance sense; in particular, exact-label recovery is a distinct
identification task, particularly for non-injective mappings.
Rather than introducing four nearly identical definitions, we formulate below
the computational and information-theoretic settings in terms of the success
event $\mathsf{Succ}(x,x')$, allowing either recovery objective to be adopted.
Unless stated otherwise, throughout this section the classical input $x$ is
assumed to be sampled uniformly from the finite domain $\mathbb X$.
}

\begin{definition}[Computational classical-to-quantum one-wayness]
\label{def1}
The mapping ${\mathbb X}\mapsto\mathbb S$ exhibits computational  one-wayness with respect to a specified recovery objective if 
the following conditions hold:
\begin{enumerate}
    \item \textbf{Efficient generation.} There exists a QPT generator $\mathcal{G}$, which on input $x\in \mathbb{X}$ returns a quantum state $\rho_x\in\mathbb{S}$.
    \item \textbf{Hardness of inversion.} For every uniform QPT adversary\footnote{By ``uniform QPT adversary'' we mean a single QPT algorithm
that works for all input lengths $n$, where the security parameter is given in unary notation as input $1^n=11\ldots 1$ (so that the length of the desired output is known). The security parameter $n$ determines the size of $\mathbb{X}$, the dimension $d$, and the allowed number of copies $c$.} $\mathcal{A}$ we have
   \[
\Pr_{x \leftarrow \mathbb{X},\; x' \leftarrow \mathcal{A}(\rho_x^{\otimes c},1^n)}
\Big[
\mathsf{Succ}(x,x')
\Big]
\le \mathrm{negl}(n).
\]
    The probability is taken over the uniformly random choice of $x\in\mathbb{X}$, as well as over the internal randomness and measurements of $\mathcal{A}$.
\end{enumerate}
\end{definition}

Definition~\ref{def1} is computational in nature, because security is defined with respect to an efficient (QPT) adversary, and inversion is assumed to be computationally infeasible despite being physically allowed in principle. We adopt the uniform QPT model throughout this review primarily for simplicity and accessibility. In the non-uniform setting, the adversary is additionally allowed to receive polynomial-size advice that may depend on the security parameter, leading to a formally stronger security notion. Most constructions discussed in this review can be formulated in either framework, and the conceptual distinctions between computational and information-theoretic quantum one-wayness remain unchanged. Since our primary goal is to provide an introduction to quantum one-way primitives for a broad readership, we restrict attention to the uniform model unless explicitly stated otherwise. More generally, one may consider $c(n)$ copies, where $c$ is any polynomially bounded function of the security parameter $n$. Security against polynomially many copies immediately implies security against any smaller number of copies, while the special case $c=1$ corresponds to a single-copy challenge.

A different notion of quantum one-wayness arises when the restriction on the adversary's computational power is removed. In this information-theoretic setting, one seeks security against arbitrary physically allowed quantum operations. In practice, such security can generally be expected only when the adversary is provided with a limited number of copies of the challenge state. The resulting one-wayness then arises from fundamental principles of quantum mechanics, including the no-cloning theorem, the Holevo bound, and limitations on the distinguishability of non-orthogonal quantum states.
For instance, in view of the Holevo bound, the amount of classical information that an adversary can extract from $c$ copies of a $d$-dimensional quantum state is at most $c\log_2(d)$ bits. Since specifying the index $x\in\mathbb{X}$ requires $n$ bits, if $n\gg c\log_2(d)$ then the adversary is information-theoretically constrained and cannot, in general, extract enough classical information to uniquely identify $x$. This rather simple information-theoretic limitation provides intuition for exact inversion hardness, and shows that the parameters $n$, $c$, and $d$ cannot be chosen arbitrarily. In particular, the more copies become available to the adversary, the more classical information can be extracted from them. Thus a mapping $\mathbb{X}\mapsto\mathbb{S}$ may exhibit information-theoretic one-wayness only up to a limited number of copies, which motivates the following definition.

\rev{
\begin{definition}[Bounded-copy information-theoretic  classical-to-quantum one-wayness]
\label{def2}
The mapping ${\mathbb X}\mapsto\mathbb S$ exhibits a bounded-copy information-theoretic
one-wayness with respect to a specified recovery objective if  the following conditions hold:
\begin{enumerate}
    \item \textbf{Efficient generation.} There exists a QPT generator $\mathcal{G}$, which on input $x\in \mathbb{X}$ returns a quantum state $\rho_x\in\mathbb{S}$.
    \item \textbf{Hardness of inversion.} For any physically allowed adversary $\mathcal A$, we have
   \[
\Pr_{x \leftarrow \mathbb{X},\; x' \leftarrow \mathcal{A}(\rho_x^{\otimes c},1^n)}
\Big[
\mathsf{Succ}(x,x')
\Big]
\le \mathrm{negl}(n).
\]
\end{enumerate}
All notation and conventions are as in Definition~\ref{def1}.
\end{definition}
}

Although Definitions~\ref{def1} and~\ref{def2} have a similar mathematical structure,
they correspond to fundamentally different adversarial models. 
Definition~\ref{def1} assumes computationally bounded (QPT) adversaries, whereas Definition~\ref{def2}
requires security against arbitrary physically allowed quantum operations. 
That is, even if an adversary can perform any quantum operation and
measurement allowed by quantum physics, it should not be able to output a
label $x'$ satisfying the chosen recovery objective with non-negligible
probability.
The most general physically allowed quantum operation is described by a completely positive trace-preserving (CPTP) map. Therefore, the adversary $\mathcal A$ appearing in Definition~\ref{def2} may be regarded as an arbitrary CPTP map.  
On the other hand, the computational formulation of Definition \ref{def1} extends the scope of quantum one-wayness beyond the regime in which information-theoretic security can be established. In particular, information-theoretic one-wayness is typically accompanied by restrictions on the number of copies available to an adversary, whereas computational one-wayness may remain meaningful even when polynomially many copies of the output state are available. This distinction is crucial for primitives such as OWSGs, PRSGs, and EFI pairs, whose security relies on computational hardness assumptions rather than solely on quantum-information-theoretic limitations.

The distinctions between computational and information-theoretic
quantum one-wayness are not always sharp, and practical protocols may
combine elements of both.
Nevertheless, separating these notions provides a useful framework
for interpreting the diverse constructions discussed throughout this review.

Definitions~\ref{def1} and~\ref{def2}  apply to both mixed and pure states, while they are sufficiently general to cover both finite-dimensional 
(qubit- or qudit-based) constructions and continuous-variable settings. 
In the finite-dimensional case, the output states live on at most polynomially 
many qubits, which follows automatically from the assumption that $\mathcal{G}$ 
is a QPT generator. 
In continuous-variable cases (e.g.\ coherent states), one typically works 
under an energy constraint, so that the effective state family can be described 
within a suitable truncated subspace. 
\rev{
For the hardness condition to be meaningful, the relevant success
probability under the chosen recovery objective must be negligible in the
security parameter. For exact-label recovery with a uniformly sampled input,
random guessing succeeds with probability $1/|\mathbb X|$, so one typically
requires $|\mathbb X|$ to grow exponentially with $n$. For approximate
inversion, however, the random-guessing probability depends additionally on
the number and structure of labels whose associated states satisfy the
acceptance criterion, and therefore cannot in general be inferred from
$|\mathbb X|$ or $|\mathbb S|$ alone.
}

In the inversion games considered above, the challenger is assumed to have access to an efficient verification procedure for determining whether the adversary has succeeded according to the chosen inversion objective. For the exact inversion objective, verification simply amounts to checking whether the returned label satisfies $x'=x$. For the approximate inversion objective, verification requires deciding whether the quantum state associated with the returned label satisfies the prescribed acceptance criterion (for example, a fidelity threshold). The particular verification procedure therefore depends on the concrete construction under consideration.
In many theoretical constructions, the family of states
$\{\rho_x\}_{x\in\mathbb X}$ is specified explicitly, allowing the success condition to be evaluated efficiently. In particular, when approximate inversion is formulated using the fidelity criterion adopted in this review, the verifier accepts whenever
\(
F(\rho_x,\rho_{x'})\ge\tau, 
\)
and the threshold parameter $\tau$ should be understood purely as a verification condition in the inversion game. 

Unlike the classical setting, the fidelity condition does not define a canonical preimage set of labels corresponding to a 
given output state; rather, it merely specifies which guesses are accepted in the inversion game. In particular, although tensor powers of non-orthogonal states 
may become nearly orthogonal as the number of copies increases, this does not by itself imply efficient inversion when the adversary must identify 
the correct state among exponentially many candidates.
Geometric separation alone therefore does not guarantee invertibility, since successful inversion depends not only on the distinguishability 
of individual states, but also on the structure of the entire state ensemble. This subtlety regarding the interpretation of 
preimages in the quantum setting has been analyzed in detail in Ref.~\cite{KhuranaTomer2023}. 
The practical choice of the verification threshold therefore depends on the geometry of the underlying state ensemble. 
In general, the states in the set $\mathbb{S}$ may have different pairwise fidelities.
The quantity
$F_{\max}:=\max_{x\neq x'}F(\rho_x,\rho_{x'})$
provides a useful characterization of the overlap structure of the state ensemble.
Although this quantity is not part of Definitions~\ref{def1} and~\ref{def2}, in many concrete constructions it is desirable that
$F_{\max}$ is bounded away from unity.  
In this case one may choose a verification threshold $\tau$ satisfying
$F_{\max}<\tau<1$. 
This guarantees that the correct label is always accepted in the ideal noiseless setting, since
$F(\rho_x,\rho_x)=1$,
while every incorrect label is rejected because
$F(\rho_x,\rho_{x'})\le F_{\max}<\tau$ for all $x'\neq x$.
By contrast, if $F_{\max}\geq\tau$, then there exists at least one pair of
distinct labels $x\neq x'$ whose associated states satisfy the acceptance
condition.
This does not by itself imply insecurity, since the set of such labels may be negligibly small or computationally hard to find. 
It does mean, however, that the threshold no longer provides a perfect separation between correct and incorrect labels. 
The condition $F_{\max}<\tau<1$ is therefore a useful design criterion in constructions where one wants the fidelity test itself to rule out all incorrect labels in the ideal noiseless setting.

In concrete cryptographic implementations, the verifier often has access only to physical quantum states rather than their classical descriptions. In such cases, direct evaluation of the fidelity is generally impossible, and verification must instead rely on suitable quantum procedures, such as the SWAP test \cite{barenco1995elementary,Buhrman2001}, or other overlap-estimation protocols \cite{NielsenChuang2000,montanaro2016quantum}. Since these procedures are probabilistic, and their acceptance statistics depend on the overlap between the compared states, it is important that the state ensemble $\mathbb{S}$ is chosen so that distinct states are neither too similar (otherwise an adversary may return an incorrect but nearby state that is accepted with non-negligible probability) nor too dissimilar (otherwise they become readily distinguishable). This tradeoff further illustrates the role of the threshold parameter $\tau$ whenever approximate inversion is verified through the fidelity criterion.

Before proceeding to specific constructions, a few remarks are in order. 

$\bullet$ {\bf \em Remark 1.} 
\rev{ For an injective classical post-quantum OWF
$f:\mathbb X\rightarrow\{0,1\}^{m(n)}$, the mapping
$x\mapsto |f(x)\rangle$ provides a special case in which exact-label
recovery and ordinary classical inversion coincide. More generally, for a
non-injective classical OWF the appropriate success condition is recovery of
any $x'$ satisfying $f(x')=f(x)$, rather than recovery of the particular
input $x$ sampled by the challenger.
}
One may therefore wonder whether classical-to-quantum OWFs should be regarded as distinct objects. The primary motivation for studying genuinely quantum outputs, however, is not merely to generalize classical OWFs, but to exploit uniquely quantum properties of the output states. Unlike classical outputs, quantum states may be non-orthogonal, cannot be cloned, and can exhibit bounded-copy behavior. In particular, the information-theoretic notion of one-wayness introduced in Definition~\ref{def2} relies fundamentally on these quantum features. A classical output string can be copied arbitrarily and read without disturbance, whereas a quantum state may reveal only limited information under measurement and, in general, cannot be cloned.  
These features enable cryptographic functionalities such as quantum money, unclonable cryptography, quantum public-key schemes, and information-theoretic forms of one-wayness that have no direct classical counterpart. 
In  such applications, the output state itself serves as a cryptographic resource, rather than merely as a classical function value whose inversion is assumed to be hard. Thus, the interest in classical-to-quantum OWFs lies not only in computational hardness, but also in the additional cryptographic capabilities arising from the quantum nature of the output states.

$\bullet$ {\bf \em Remark 2.}
It is important to emphasize that approximate inversion of a single invocation of a QOWF does not necessarily imply a successful attack on a cryptographic protocol employing that QOWF. In many applications, the primitive is invoked repeatedly, and the adversary must succeed on many independent instances before the protocol is compromised. Consequently, even if approximate inversion succeeds with non-negligible probability for an individual invocation, the overall success probability of an attack may still be made arbitrarily small through the design of the higher-level protocol. The operational significance of approximate inversion therefore depends on the particular cryptographic application and should not be inferred solely from the security of an isolated invocation of the underlying QOWF.

Having introduced the notion of a classical-to-quantum OWF in a general theoretical framework, we now discuss representative constructions that may be viewed as concrete realizations of the above definitions. Some of these constructions derive their one-wayness primarily from information-theoretic limitations associated with quantum states, while others rely on computational hardness assumptions.
Unless stated otherwise, from now on the term QOWF refers to a classical-to-quantum OWF.

\subsubsection{Construction 1: Conjugate coding.}
\label{sec:const1}

Let ${\bm a},{\bm b}\in\{0,1\}^n$ be uniformly random and independent
classical bit strings. Consider two mutually unbiased bases of a qubit,
\[
{\cal B}_0:=\{\ket{0}_z,\ket{1}_z\},
\qquad
{\cal B}_1:=\{\ket{0}_x,\ket{1}_x\},
\]
where
\[
\ket{j}_x=
\frac{\ket{0}_z+(-1)^j\ket{1}_z}{\sqrt{2}}.
\]
For each position $i\in[1,n]$, the bit $a_i$ is encoded in the
corresponding eigenstate of the basis ${\cal B}_{b_i}$, where
$b_i\in\{0,1\}$ is chosen uniformly at random. For a fixed realization
of the basis string ${\bm b}$, the resulting pure state is
\begin{equation}
|\psi_{\bm a}\rangle_{\bm b}
=
\bigotimes_{i=1}^n |a_i\rangle_{b_i}.
\label{CCmap:eq}
\end{equation}

\rev{
Here ${\bm a}$ is the classical input to be recovered, whereas ${\bm b}$
is regarded as internal randomness of the state-preparation procedure.
Accordingly, the classical-to-quantum mapping relevant to the inversion
game is
\(
{\bm a}\mapsto\rho_{\bm a},
\)
where
\begin{equation}
\rho_{\bm a}
=
2^{-n}
\sum_{{\bm b}\in\{0,1\}^n}
|\psi_{\bm a}\rangle_{\bm b}
\langle\psi_{\bm a}|_{\bm b}.
\label{CCmixed:eq}
\end{equation}
}

\rev{
This preparation procedure lies at the core of QKD protocols and several
other quantum-cryptographic primitives
\cite{Gisin2002QuantumCryptography,BroadbentSchaffner2016,
BozzioCrepeauWalldenWalther2024,Wiesner1983,
BennettBrassard1984BB84,Bozzio2018QuantumMoney}.
It is efficient, since it requires the preparation of only $n$
single-qubit states, with the bases chosen according to ${\bm b}$ and
the encoded values according to ${\bm a}$.
In the usual conjugate-coding setting underlying QKD, each execution
uses freshly and independently chosen strings ${\bm a}$ and ${\bm b}$,
and the adversary receives only one realization of the corresponding
$n$-qubit state. Thus, in the terminology of Definitions~\ref{def1} and~\ref{def2}, the
natural QKD-like setting corresponds to $c=1$. If one considers instead
$c>1$ copies of the same challenge $\rho_{\bm a}$, as allowed in
Definitions~\ref{def1} and~\ref{def2}, the encoded string ${\bm a}$ is
held fixed while the preparation randomness ${\bm b}$ is independently
resampled for each copy. The latter reflects the fresh basis choice
characteristic of independent executions of conjugate-coding protocols,
although in QKD the encoded string ${\bm a}$ itself is also refreshed
between executions.
}

From the adversary's perspective, and before conditioning on the value
of the encoded bit, each qubit is uniformly distributed over the four
states
\[
\{\ket{0}_z,\ket{1}_z,\ket{0}_x,\ket{1}_x\}.
\]
Its unconditional average state is therefore
\[
\bar{\rho}
=
\frac{1}{4}
\sum_{a_i,b_i}
|a_i\rangle_{b_i}\langle a_i|_{b_i}
=
\frac{\mathbb I_2}{2},
\]
where ${\mathbb I}_2$ is the two-dimensional identity operator.
According to the Holevo bound, an adversary can extract at most $n$
bits of classical information from the $n$ qubits. This observation
alone does not imply hardness of label recovery, since ${\bm a}$ itself
contains $n$ bits. The relevant limitation arises from the fact that
the states encoding the two possible values of each bit are
non-orthogonal.

Because the individual positions are prepared independently, it is
sufficient to consider the recovery of a single bit $a_i$. Conditioned
on its value, the corresponding mixed states are
\[
\rho_0
=
\frac{1}{2}
\left(
\ket{0_z}\!\bra{0_z}
+
\ket{0_x}\!\bra{0_x}
\right),
\qquad
\rho_1
=
\frac{1}{2}
\left(
\ket{1_z}\!\bra{1_z}
+
\ket{1_x}\!\bra{1_x}
\right).
\]
The optimal strategy for identifying $a_i$ is the Helstrom
measurement, which in this case is a projective two-outcome measurement
in the Breidbart basis, i.e., a basis rotated by $\pi/8$ with respect
to both ${\cal B}_0$ and ${\cal B}_1$. The corresponding optimal
success probability is
\[
p_{\mathrm{opt}}^{(1)}
=
\frac{1}{2}
\left(
1+\frac{1}{\sqrt{2}}
\right)
=
\cos^2\!\left(\frac{\pi}{8}\right)
\approx0.8536.
\]

For the single-copy $n$-qubit challenge state, measuring each qubit
independently therefore gives
\[
p_{\mathrm{opt}}^{(n)}
=
\left[
\frac{1}{2}
\left(
1+\frac{1}{\sqrt{2}}
\right)
\right]^n
=
2^{-\nu n},
\]
where
\[
\nu
=
1-\log_2\!\left(1+\frac{1}{\sqrt{2}}\right)
\simeq0.2284.
\]
For this product ensemble and the task of recovering ${\bm a}$ exactly,
the optimal single-qubit measurements yield the optimal probability of
recovering the entire string. Since the success probability per bit is
strictly smaller than unity, the probability of exact-label recovery
decreases exponentially with $n$. The hardness arises from the
non-orthogonality of the quantum states and the resulting impossibility
of perfect state discrimination.

The Holevo bound constrains only the \emph{average accessible
information} obtainable from the ensemble and does not directly
determine the probability of recovering ${\bm a}$ perfectly. In
particular, even when the accessible information scales as $O(n)$, the
optimal probability of identifying the complete string may still
decrease exponentially with $n$ (see Appendix~\ref{appendix2} for a
summary of asymptotic notation).

\rev{
The above analysis establishes information-theoretic hardness of
\emph{exact-label recovery} in the single-copy setting: the probability
of recovering the complete encoded string ${\bm a}$ decreases
exponentially with $n$. As emphasized above,
exact-label recovery is a distinct security objective and should not,
by itself, be identified with classical preimage resistance.
}

\rev{
For completeness, for $c>1$ the states to be discriminated are
$\rho_0^{\otimes c}$ and $\rho_1^{\otimes c}$. The single-copy states
can be written as
\[
\rho_0=\frac{\mathbb I_2}{2}+\frac{\hat {X}+\hat{Z}}{4},
\qquad
\rho_1=\frac{\mathbb I_2}{2}-\frac{\hat {X}+\hat {Z}}{4},
\]
where $\hat {X}$, and $\hat {Z}$ denote the Pauli operators.
Hence, they commute and are simultaneously diagonal in the eigenbasis
of $\hat {X}+\hat {Z}$, which is the Breidbart basis. Independent Breidbart
measurements followed by the optimal classical decision rule therefore
achieve the optimal discrimination probability, and collective quantum
measurements provide no advantage. Since the $c$ measurement outcomes
are conditionally independent, the optimal classical maximum-likelihood
decision is majority voting, with a random decision in the case of a tie
(which can occur only for even $c$). 
For every  finite $c$, the corresponding per-bit success probability
remains strictly smaller than unity, and hence the probability of recovering
the complete $n$-bit string still decreases exponentially with $n$.
}

\subsubsection{Construction 2: Quantum-fingerprint states.}
\label{sec:const2}

Another concrete realization of QOWFs is provided by the \emph{quantum fingerprint} construction of 
Buhrman \emph{et al.}~\cite{Buhrman2001}.
Let $\mathbb{X}=\{0,1\}^n$ and fix an efficiently computable binary error-correcting code 
$E:\{0,1\}^n\to\{0,1\}^d$ with 
$d=\zeta\cdot n$ (for some positive constant $\zeta$), where for any two distinct codewords we have   the Hamming distance 
$\Delta\ge (1-\delta) d$, for some positive $0<\delta<1$\footnote{In the case of Justesen codes, we may choose any $\zeta >2$ and we will have $\delta < 0.9 + 1/(15\zeta)$, for sufficiently large $n$.}. 
For each $x\in\mathbb{X}$ define the pure bipartite \emph{fingerprint state} 
\begin{eqnarray}
|\psi_x\rangle = \frac{1}{\sqrt{d}}\sum_{i=1}^{d} |i\rangle\,|E_i(x)\rangle. 
\label{fingerprint:eq1}
\end{eqnarray}
The first register pertains to a $d$-dimensional system, where 
$q(n)=\lceil\log_2 d\rceil=O(\log n)$ qubits. The second register 
explicitly encodes each bit $E_i(x) \in \{0,1\}$ of the codeword $E(x)\in\{0,1\}^d$. 
The fingerprint states can be also expressed in a phase-encoded form  as follows 
\begin{eqnarray}
|\psi_x\rangle=\frac{1}{\sqrt d}\sum_{i=1}^{d}(-1)^{E_i(x)}|i\rangle.
\label{fingerprint:eq2}
\end{eqnarray}
This expression pertains to a single register and it is equivalent to Eq. (\ref{fingerprint:eq1}), up to a simple isometry that maps $|i\rangle|b\rangle\mapsto(-1)^b|i\rangle$. 
%Clearly, we have $2^n$ different fingerprint states, and the mapping $x\mapsto\ket{\psi_x}$ is one-to-one. 

Taking the inner product between two distinct states we obtain
 $\left |\langle\psi_x|\psi_{x'}\rangle \right | \le \delta$, and hence the maximum fidelity between distinct states is 
$F_{\max}=\delta^2$. 
Choosing a constant threshold $\tau$ such that 
$\delta^2<\tau<1$ ensures that no incorrect label $x'\neq x$ can satisfy the 
acceptance condition $F(\rho_x,\rho_{x'})\ge\tau$, while the correct label always does. 
Hence, for any verification threshold satisfying $\delta^2<\tau<1$, the approximate inversion objective is equivalent to the exact inversion objective. Consequently, the fingerprint construction naturally satisfies both inversion objectives introduced above.

For any fixed constants $(\tau,c)$, and given that the Hilbert-space dimension
of our encoding grows linearly as $d = \zeta n$, the Holevo bound implies that
the total amount of accessible classical information from $c$ copies of the
state is at most  $c \log d = c\log(\zeta n) = O(c\log n)$. 
\rev{
Hence, if $c$ is constant, or more generally if
$c(n)=o\!\left(\frac{n}{\log n}\right)$, then
$c(n)\log d=o(n)$, so the accessible classical information is
asymptotically much smaller than the $n$ bits required to specify the input.
This provides an information-theoretic obstruction to recovering the
original string $x$ and motivates viewing the mapping
$x\mapsto|\psi_x\rangle$ as a bounded-copy information-theoretic QOWF.
}

If $c(n)$ grows polynomially with $n$, however, the quantity $c(n)\log d$ may be
of order $n$ or larger. In this case, the Holevo bound no longer yields a
meaningful information-theoretic restriction on the inversion of the mapping, and one-wayness
must then be understood in the \emph{computational} sense of our definition~\ref{def1}
rather than as an information-theoretic consequence. Whether the fingerprint
mapping remains computationally hard to invert for polynomially many copies
$c(n)$ is not clear, and we are not aware of a proof either way. 
Consequently, quantum fingerprint states provide a canonical example of a bounded-copy information-theoretic QOWF with respect to both the exact and approximate inversion objectives, and have served as the basis for numerous quantum cryptographic applications.

It is worth emphasizing that the significance of the fingerprint-state construction lies in its 
ability to embed an exponentially large set of classical strings, $|\mathbb{X}| = 2^n$, into a Hilbert space of only polynomial dimension $d = O(n)$, 
while maintaining pairwise overlaps that are small and bounded away from one.  
In other words, one can efficiently prepare exponentially many $(2^n)$ distinct quantum states 
that remain well separated in  Hilbert space. This near-orthogonality allows, in principle, their discrimination with 
 small error given sufficiently many copies, but when only a limited number of copies are available, 
the probability of correct discrimination remains bounded away from one, 
which effectively limits the information an adversary can extract and underlies 
the one-wayness of the mapping $x\mapsto|\psi_x\rangle$.

Despite its conceptual elegance, the fingerprint-state construction also has certain limitations 
when viewed as a candidate for a QOWF. 
First,  the pairwise overlaps between distinct states are constant and do not decrease with~$n$, 
which means that with sufficiently many copies the states can in principle be distinguished almost perfectly. 
Second, practical implementation of quantum fingerprints for large values of $n \gg 1$, faces significant experimental challenges related mainly to maintaining coherence of the quantum states during preparation, transmission, and storage.

Quantum fingerprint states were first exploited by Buhrman et al. \cite{Buhrman2001} as a means of extending classical fingerprinting to the quantum setting. Fingerprinting is a cryptographic primitive that allows two honest parties to determine whether they hold identical or different binary strings by sending shorter fingerprints to a trusted authority. In the classical setting, fingerprints cannot be exponentially shorter than the original $n-$bit strings unless the parties have access to sources of correlated randomness. 
In particular, without shared randomness,  the length of classical fingerprints must scale at least proportionally to $\sqrt{n}$ in the standard fingerprinting model.
Buhrman et al. showed that in the quantum setting the problem can be solved using $O(\log_2 n)-$qubit fingerprints, yielding an exponential improvement over the classical bound.

To the best of our knowledge,  Gottesman and  Chuang \cite{gottesman2001} were the first to observe that quantum fingerprint states possess properties consistent with information-theoretic QOWFs. They subsequently proposed an information-theoretically secure quantum digital signature scheme inspired by the classical Lamport–Diffie construction, in which the quantum states $\{\ket{\psi_x}\}$ serve as public keys and $x \in \mathbb{X}$ acts as the private key.
The scheme of Gottesman and Chuang is not feasible with current technology, primarily because it relies on SWAP tests and requires long-term quantum memory, as the public quantum keys must be stored indefinitely. Relaxing these requirements and enabling experimental realizations has been the subject of extensive research; the interested reader may consult relevant review articles and references therein \cite{AmiriAndersson2015,BozzioCrepeauWalldenWalther2024}.
It is worth noting, however, that the most prominent experimental demonstrations employ a different type of QOWF, based on encoding classical information into coherent states of light (see Construction 4 below). Moreover, many such implementations share a common structural constraint: during an initial distribution phase, the signer distributes signature material corresponding to all possible future messages.
This requirement originates from the underlying security models, which rely on message-specific quantum states (or classical data derived from them) that must be pre-shared and stored by the recipients. Consequently, these implementations typically operate in a one-time or few-time signature regime, where the set of messages to be signed is fixed in advance.
Hybrid schemes that combine quantum states with classical primitives, such as strong-universal hash functions, relax this requirement within the information-theoretic framework and aim to improve scalability. Alternatively, one may relax unconditional security, e.g., by considering computationally hard QOWFs.

\subsubsection{Construction 3: Single-qubit states.}
\label{sec:const3}

A considerably simpler construction was introduced in \cite{NikolopoulosPRA2008} and pertains to single-qubit rotations.  
We will adopt here a slightly different mapping, but the main idea remains the same. 
Consider a random integer $x$ uniformly distributed over the set 
${\mathbb X} = \{0,1,\ldots, 2^n-1\}$ for some fixed positive integer $n\gg 1$. 
For a given randomly chosen $x\in{\mathbb X}$, we prepare the qubit state 
\bea
\ket{\psi_x(\theta_n)}=\frac{1}{\sqrt{2}}\left (\ket{0}+e^{{\rm i}x\theta_n}\ket{1}\right ),
\label{1s}
\eea
where $(\ket{0}+\ket{1})/\sqrt{2}$ is rotated  by $x\theta_n$ around the $z$-axis with $\theta_n:=\pi/2^{n-1}$.
Hence, for some fixed $n\in\PosInt$, the output of the QOWF  
pertains to the set of qubit states 
\[{\mathbb S}=\{\ket{\psi_x(\theta_n)}| x\in{\mathbb X}, \theta_n=\pi/2^{n-1}\}.\]
As $n$ increases, the number of states in the ensemble grows exponentially, 
while the angular separation between neighboring states decreases. 
In particular, for $n\gg 1$ the fidelity between neighboring states is  
\[
F_{\max} = |\langle\psi_x(\theta_n)|\psi_{x+1}(\theta_n)\rangle|^2 =\cos^2\left (\frac{\pi}{2^n}\right ).
\]
Hence, unlike the fingerprint-state construction, the maximum pairwise fidelity between distinct states approaches unity as $n$ increases, 
meaning that neighboring states on the Bloch sphere become arbitrarily close. 
Consequently, the parameters $n$ and $\tau$ cannot be chosen independently. 
Increasing $n$ improves the information-theoretic hiding of the label $x$, but also increases the overlap between distinct states, 
which may complicate verification procedures based on state comparison. 
The appropriate choice of these parameters therefore depends on the particular cryptographic application and verification strategy under consideration.

For a given pair of integers $\{n,x\}$, the function $x\mapsto\ket{\psi_x(\theta_n)}$ 
is easy to compute since it involves single-qubit rotations only. In particular, 
it is known that any single-qubit operation can be simulated to an arbitrary accuracy 
$\epsilon>0$, by a quantum algorithm involving a universal set of gates 
(i.e., Hadamard, phase, controlled-NOT, and $\pi/8$ gates) \cite{NielsenChuang2000}. 
Moreover, this simulation is efficient since its implementation 
requires an overhead of resources that scales polynomially with $\log(\epsilon^{-1})$. 
The states (\ref{1s}) are phase-encoded qubit states lying on the equator of the Bloch sphere. 
Such states can be prepared and manipulated using standard techniques developed for phase-encoded QKD systems, 
where information is encoded in the relative phase between two optical modes or time bins. 
Consequently, the experimental requirements for realizing the mapping $x\mapsto\ket{\psi_x(\theta_n)}$ are closely related to well-established QKD technology.

Inversion of the map $x\mapsto\ket{\psi_x(\theta_n)}$ means to recover $x$ 
from a given qubit state $\ket{\psi_x(\theta_n)}$ chosen at random from the set 
of all possible qubit states
${\mathbb S}_n:=\{\ket{\psi_x(\theta_n)}~:~x\in\mathbb{X}\}$.  
Inversion is closely related to the problem of discriminating among the non-orthogonal states in $\mathbb S_n$, 
and it is information-theoretically constrained by the Holevo bound \cite{holevo1973bounds}, 
since the $n$-bit label $x$ is encoded into a single qubit. 
\rev{For constant $c\ll n$, or more generally for $c(n)=o(n)$, the elementary
Holevo bound already limits the amount of accessible classical information.
As shown below, however, the symmetry of the $c$ identical copies yields a
considerably stronger bound.
}

The above dimensional argument that treats $c$  two-dimensional systems as an unconstrained 
$2^c$-dimensional Hilbert space, ignores the symmetry, and the structure of the ensemble accessible by the adversary.
In the actual inversion problem, an adversary is given $c$ identical copies 
of a randomly chosen state $\ket{\psi_x}$. The corresponding state 
\begin{equation}
\rho_x^{(c)} = (\ket{\psi_x}\bra{\psi_x})^{\otimes c}
\end{equation}
is invariant under any permutation of the $c$ copies, and thus it lies 
in the symmetric subspace of $(\mathbb{C}^2)^{\otimes c}$, whose dimension is (see appendix \ref{appendix3})
\begin{equation}
D_{\mathrm{sym}}(c,d=2) = c+1.
\label{Dim_sym2:eq}
\end{equation}

\rev{
This dimensional restriction also gives a direct bound on the probability
of exact-label recovery. For a uniformly distributed ensemble of
$2^n$ pure states supported on a $D$-dimensional subspace, the average
success probability of any measurement that attempts to identify the
preparation label satisfies
\[
P_{\rm succ}^{\rm exact}\leq \frac{D}{2^n}.
\]
Since $c$ identical copies of the present single-qubit states are supported
on the symmetric subspace of dimension $D=c+1$, it follows that
\[
P_{\rm succ}^{\rm exact}\leq \frac{c+1}{2^n}.
\]
Thus exact-label recovery remains negligible for polynomially bounded
$c$. Importantly, this bound follows essentially from the dimensional
compression of exponentially many classical labels into the symmetric
subspace and does not rely on special properties of the particular
single-qubit ensemble considered here.
}

Given that from the adversary's point of view the pure state  $\ket{\psi_x}$ is chosen at random from ${\mathbb S}$ with probability $p_x$, 
the upper bound to the accessible information is determined by the von Neumann entropy of the average state 
$\bar{\rho}^{(c)}  = \sum_x p_x \rho_x^{(c)} $. As shown in appendix \ref{appendix4}, the average state $\bar{\rho}^{(c)}$ is diagonal in the Dicke basis.   
By the Holevo bound, the accessible information obeys
\begin{equation}
I_{\mathrm{acc}} \le S(\bar{\rho}^{(c)} ) 
\le \log_2 \big( \mathrm{rank}(\bar{\rho}^{(c)} ) \big),
\end{equation}
while 
\begin{equation}
\mathrm{rank}(\bar{\rho}^{(c)} ) \le D_{\mathrm{sym}}(c,d=2), 
\end{equation}
and equality holds for sufficiently rich and balanced (uniform) ensembles that span the entire symmetric subspace. 
Consequently, whenever $n\gg \log_2(c+1)$, the accessible information is far smaller than the number of bits required to specify $x$, 
providing strong information-theoretic evidence for one-wayness of the mapping.
More refined bounds on the accessible information can be obtained by considering explicitly 
the states that comprise the ensemble and computing the corresponding entropy $S(\bar{\rho}^{(c)} )$ \cite{SeyfarthNikolopoulosAlberPRA2012}.

\rev{
The above analysis establishes information-theoretic hardness of
\emph{exact-label recovery}, since the adversary is required to identify
the particular classical label $x$. This result should not by itself be interpreted as establishing
preimage-resistant one-wayness in the classical sense. 
By contrast, the approximate inversion objective is substantially different for this construction. Indeed, the maximum fidelity between distinct states satisfies
$F_{\max}\rightarrow1$ as $n$ increases, 
so no fixed verification threshold $\tau<1$ can separate correct from incorrect labels. Consequently, the information-theoretic analysis presented here does not by itself establish hardness with respect to the approximate inversion objective.
}

Thanks to its simplicity and its close relation to QKD technology, the mapping $x \mapsto \ket{\psi_x(\theta_n)}$ with $\ket{\psi_x(\theta_n)}$ given by Eq. (\ref{1s}),
 has also been exploited in the design of 
quantum public-key encryption schemes \cite{NikolopoulosPRA2008}. The corresponding security analysis involves attack models that do not typically arise in QKD, 
such as forward-search, chosen-plaintext, and chosen-ciphertext attacks \cite{NikolopoulosPRA2008,NikolopoulosIoannouPRA2009,SeyfarthNikolopoulosAlberPRA2012}. In analogy to the Gottesman–Chuang quantum signature scheme, in quantum public-key encryption the value $x \in \mathbb{X}$ serves as the private secret key, while $\ket{\psi_x(\theta_n)}$ functions as the quantum public key, which is distributed in a limited number of copies. The quantum ciphertext is obtained by encoding the plaintext onto the quantum public-key state.  Security relies on distributing only a limited number of copies of each quantum public key, since unrestricted copying would eliminate the information-theoretic advantage provided by the quantum encoding. Similarly to various quantum digital signatures schemes, quantum public-key encryption requires long-term quantum memories for the storage of these quantum public keys. 
%%A quantum public-key encryption scheme, which relies on quantum walks rather than single-qubit rotations, has also been proposed by Vlachou {\em et al.} \cite{Vlachou15}.
\rev{
The role of approximate inversion in concrete cryptographic protocols based on this construction depends on the protocol itself. As discussed in Remark~2 above, a non-negligible success probability for approximate inversion of a single instance does not necessarily imply a successful attack, since many applications invoke the QOWF repeatedly, causing the overall cheating probability to decrease rapidly with the number of independent instances.
}

\subsubsection{Construction 4: Phase-encoded  coherent states}
\label{sec:const4}

 Let us consider now the mapping $x\mapsto \ket{\psi_x}$, between the set of integers ${\mathbb X}:=\{0, 1, \ldots N-1\}$ and the 
 set of single-mode coherent states 
\begin{subequations}
\label{CoheSet:Eqs}
 \begin{eqnarray}
{\mathbb S}_N \equiv \left \{\ket{\psi_x}:=\ket{\sqrt{\mu}e^{{\rm i} x \delta\varphi}}~:~\delta\varphi:=\frac{2\pi}N,~ x \in {\mathbb X}\right \},
\label{set_coh:eq}
\end{eqnarray} 
for even $N>4$, where $\mu$ is the mean number of bosons and $\ket{\psi_x}$ is eigenstate of the bosonic annihilation operator $\hat{a}$ i.e., 
$\hat{a}\ket{\psi_x} = \psi_x \ket{\psi_x}$. The coherent state $\ket{\psi_x}$ can be expanded in terms of the Fock states 
$\{\ket{n}\}$ as follows 
\bea
\ket{\psi_x}:=e^{-\mu/2} \sum_{n=0}^\infty \frac{\psi_x^n}{\sqrt{n!}}\ket{n}.
\label{alpha_fock:exp}
\eea
In phase space, all the coherent states in ${\mathbb S}_N$ have the same mean number of bosons $\mu$, but their phases are distributed around the circle at regular intervals of $\delta\varphi$.
%Let $\hat{\Pi}$ denote the projector onto the subspace spanned by the states in ${\mathbb S}_N$. 
The unitary transformation which maps each state onto its successor 
is the elementary phase-shift operator $\hat{\cal U} = e^{{\rm i} 2\pi \hat{a}^\dag \hat{a}/N} =e^{{\rm i} \delta\varphi \hat{a}^\dag \hat{a}}$ ,
where $\hat{a}^\dag$ is the bosonic creation operator. 
\end{subequations}
The overlap between neighboring coherent states is readily obtained from the well-known inner product of coherent states. For two adjacent labels we have
\[
|\langle \psi_x|\psi_{x+1}\rangle|
=
\exp\!\left[-2\mu\sin^2\!\left(\frac{\pi}{N}\right)\right]
\simeq
\exp\!\left(-\frac{\mu\,\delta\varphi^2}{2}\right),
\]
where the approximation holds for large $N$. Consequently, the maximum pairwise fidelity between distinct states is
\[
F_{\max}
=
\exp\!\left[-4\mu\sin^2\!\left(\frac{\pi}{N}\right)\right]
\simeq
\exp\!\left(-\mu\,\delta\varphi^2\right).
\]
Thus, for fixed $\mu$, increasing $N$ improves the information-theoretic hiding of the label $x$ by enlarging the set of possible states, but at the same time 
it increases the overlap between neighboring states. As in Construction~3, the parameters $N$, $\mu$, and the verification threshold $\tau$ cannot be chosen independently, and their optimal choice depends on the requirements of the cryptographic application under consideration.

Coherent states are standard information carriers in various quantum
communication tasks, including QKD protocols. In essentially all
practical quantum-communication implementations known to date, the
underlying bosonic systems are optical modes and the corresponding
bosonic excitations are photons. Consequently, coherent states of light
can be generated, manipulated, transmitted, and detected using mature
optical technology.
In particular, it is easy to perform the mapping $x\mapsto \ket{\psi_x}$  for a given integer $x\in{\mathbb X}$, 
using standard phase modulators. The remaining question  therefore is the conditions under which inversion of the mapping becomes information-theoretically hard.

As discussed in section \ref{sec:const3}, for $c$ i.i.d copies of $d-$dimensional quantum state, the accessible information is bounded by the logarithm of the {\em finite-dimensional} symmetric subspace of the 
Hilbert space. For coherent state $\ket{\psi_x}$, the density operator $\rho_x^{(c)} =\ket{\psi_x}\!\bra{\psi_x}^{\otimes c}$ is still 
symmetric with respect to any permutation of the $c$ copies. 
However, the single-mode Hilbert space $\mathcal{H}_\infty$ is
infinite-dimensional, and thus the symmetric subspace of
$(\mathcal{H}_\infty)^{\otimes c}$ is infinite-dimensional as well. 
 So,  in this case we cannot obtain a finite logarithmic bound, as in the case of finite systems. 
Employing the subadditivity of the von Neumann entropy we have  \cite{NielsenChuang2000}
\[ 
S(\bar{\rho}^{(c)})\leq c S(\bar{\rho}^{(1)})\leq c S_{\rm th}(\mu),
\]
where $\bar{\rho}^{(1)} = \sum_x p_x \ket{\psi_x}\!\bra{\psi_x}$
\footnote{Note that tracing out all but one copies in $\bar{\rho}^{(c)}$, we recover $\bar{\rho}^{(1)}$. Moreover there  exist cases where $\bar{\rho}^{(1)}$ is very close to diagonal (see appendix \ref{appendix5}).}, and we have used the fact that 
the entropy of $\bar{\rho}^{(1)} $ is bounded by the entropy  of the thermal state with the same mean number of photons given by 
\[S_{\rm th}(\mu) := (\mu+1)\log_2(\mu+1) - \mu \log_2(\mu).\]
In the relation above, the first equality holds only if $\bar{\rho}^{(c)} = (\bar{\rho}^{(1)})^{\otimes c}$. 
So, by virtue of the Holevo bound, in the case of coherent states,  one expects inversion of  the mapping $x\mapsto\ket{\psi_x}$ to become information-theoretically hard when 
\bea
\log_2(N)\gg c \left [ (\mu+1)\log_2(\mu+1) - \mu \log_2(\mu) \right ]. 
\label{owf_hoh:ineq}
\eea

We see that the right-hand side of Eq.~(\ref{owf_hoh:ineq}) scales linearly with the number of copies $c$, in contrast to the logarithmic dependence obtained for qubit ensembles. This difference originates from the infinite-dimensional Hilbert space of coherent states. From a cryptographic perspective, the linear scaling is significantly more restrictive, since the amount of accessible information increases rapidly with the number of copies available to an adversary. Consequently, satisfying Eq.~(\ref{owf_hoh:ineq}) may require very large values of $N$ when many copies are available. This can lead to practical complications in implementations, since larger values of $N$ generally imply a larger set of phase-encoded states and increased demands on state preparation, modulation, and verification. It is therefore important to investigate whether, and under what conditions, the logarithmic dependence on the number of copies $c$, characteristic of finite-dimensional qubit ensembles, can also be recovered in the case of coherent-state QOWFs. As we show below, this is indeed possible in the weak-photon regime $\sqrt{c}\mu\ll1$, where the coherent-state ensemble behaves effectively as a qubit ensemble.

In view of the expansion (\ref{alpha_fock:exp}), the single-mode coherent state for a randomly chosen $x\in{\mathbb X}$ can be written as 
$|\psi_x\rangle = |\tilde{\phi}_x\rangle+|\tilde{\Phi}_x\rangle$, 
where $|\tilde{\phi}_x\rangle := e^{-\mu/2}\big(\ket{0}+\psi_x\ket{1}\big)$ pertains to vacuum state and the single-photon sector only, while higher excitations with $n\geq 2$ 
are included in $|\tilde{\Phi}_x\rangle$. 
When the mean photon number  $\mu$ is very small (i.e., $\mu \ll 1$),  the Fock states $\ket{0}$ and $\ket{1}$ 
are  populated with probability $ \langle \tilde{\phi}_x|\tilde{\phi}_x\rangle  = e^{-\mu}(1+\mu) \approx (1-\mu^2/2)$, while the population of the states with higher $n\geq 2$ is equal to 
$\langle \tilde{\Phi}_x|\tilde{\Phi}_x\rangle = 1-e^{-\mu}(1+\mu) \approx \mu^2/2$. Hence, for $\mu\ll 1$ the coherent state $\ket{\psi_x}$ is well approximated by 
$|\tilde{\phi}_x\rangle = e^{-\mu/2}\big(\ket{0}+\psi_x\ket{1}\big)$, where there is a loss of normalization $O(\mu^2)$, 
because we have ignored the multiphoton sector $|\tilde{\Phi}_x\rangle$. 
In other words, the single-mode coherent state behaves as an \emph{effective} qubit 
living in the span of $\{\ket{0},\ket{1}\}$, and whose ``excited'' component amplitude
is proportional to $\psi_x$. Normalizing the state $|\tilde{\phi}_x\rangle$ we obtain 
\bea
|\phi_x\rangle = \frac{\ket{0}+\psi_x\ket{1}}{\sqrt{1+\mu}},
\label{eq:coh_qubit}
\eea
which yields for the trace norm between $c$ copies of the truncated state $\sigma_x^{(c)} = (\ket{\phi_x}\bra{\phi_x})^{\otimes c}$, and  $c$ copies of the actual state 
$\rho_x^{(c)}$ 
\bea
\left |\left |\rho_x^{(c)} - \sigma_x^{(c)}  \right |\right |_1 
  = 2\sqrt{1-|\langle \phi_x| \psi_x\rangle|^{2c}} = 2\sqrt{1-(1+\mu)^c e^{-c\mu}} \approx \sqrt{2c}\mu.  
\eea 
The trace distance is convex and thus we obtain for the phase-averaged states 
\bea
D = \frac{1}2\left |\left |\bar{\rho}^{(c)} - \bar{\sigma}^{(c)}  \right |\right |_1 \lesssim \sqrt{c/2}\mu.  
\eea
Hence, for $\sqrt{c}\mu\ll 1$ we have  that  the phase-averaged coherent state can be represented by an effective phase averaged qubit state. 

Our task now is to obtain an upper bound on $S(\bar{\rho}^{(c)} )$. Using Audenaert–Fannes inequality \cite{Audenaert2007} we have 
\bea
\left |S(\bar{\rho}^{(c)} ) -S(\bar{\sigma}^{(c)} ) \right | \leq D\log_2(d-1)+h_2(D),
\eea
where $d = c+1$ is the dimension of the truncated space\footnote{In Audenaert–Fannes continuity bound both states have to lie in the same finite-dimensional space. 
In our case, assuming $\sqrt{c}\mu\ll 1$, the presence of multiphoton $(n\geq 2)$ components in the state $\bar{\rho}^{(c)}$ is negligible, and can be ignored, 
thereby yielding $d=c+1$ [see Eq. (\ref{Dim_sym2:eq})].}, 
and $h_2(\cdot)$ denotes the binary entropy. 
Given that  $S(\bar{\sigma}^{(c)} ) \leq \log_2[{\rm rank}(\bar{\sigma}^{(c)} )] = \log_2(c+1)$ we have  
\bea
S(\bar{\rho}^{(c)})
\lesssim
\log_2(c+1)
+
D\log_2 c
+
h_2(D),
\qquad
D\lesssim \sqrt{c/2}\,\mu.
\eea
Thus, in the regime $\sqrt{c}\mu\ll1$, the leading scaling is
approximately logarithmic in the number of copies,
\bea
S(\bar{\rho}^{(c)})\sim \log_2(c+1),
\eea
up to corrections associated with the neglected multiphoton sector.

The condition $\sqrt{c}\mu\ll1$ should be viewed as a technical requirement
for approximating the coherent-state ensemble by an effective qubit
ensemble. It is considerably stronger than the usual weak-coherent-state
condition $\mu\ll1$, since the approximation must remain valid for the
entire block of $c$ copies. In this regime the relevant contribution is effectively 
restricted to the subspace generated by the vacuum and single-photon components of each copy.
Consequently, the coherent-state ensemble
behaves similarly to the symmetric qubit ensemble, recovering the
$\log_2(c+1)$ scaling of the Holevo bound.

The present mapping shares the conceptual simplicity of the previous construction based on single-qubit rotations, 
while being less demanding experimentally since it relies on weak coherent states rather than qubit sources. 
Unlike fingerprint-state constructions, coherent-state QOWFs can be implemented using standard optical technology 
and have formed the basis of some of the most prominent experimental demonstrations of quantum digital signatures \cite{AmiriAndersson2015}. 
More recently, they have been employed in a key-exchange protocol based on the interference of weak coherent states, 
which may be viewed as a quantum analogue of the classical Diffie--Hellman key-exchange protocol \cite{NikolopoulosQDH2025}. 
In contrast to quantum public-key encryption schemes, the quantum Diffie--Hellman protocol utilizes standard QKD technology and does not require long-term quantum memory.

The above analysis establishes one-wayness with respect to the \emph{exact inversion objective}, since the adversary is required to recover the classical label $x$. By contrast, the approximate inversion objective is substantially different for this construction. As discussed above, for fixed mean photon number $\mu$, the maximum fidelity between distinct coherent states satisfies $F_{\max}\rightarrow 1$ as $N$ increases. Consequently, no fixed verification threshold $\tau<1$ can asymptotically separate correct from incorrect labels, and the information-theoretic analysis presented here does not directly establish hardness with respect to approximate inversion. Nevertheless, this does not imply that cryptographic protocols based on coherent-state QOWFs become insecure. As discussed in Remark~2,  the operational significance of approximate inversion depends on the protocol itself. 
In particular, as illustrated by the quantum Diffie–Hellman protocol of Ref.~\cite{NikolopoulosQDH2025}, cryptographic protocols may invoke the underlying QOWF repeatedly or incorporate additional verification, coding, or privacy-amplification procedures, so that the overall probability of a successful attack can be made arbitrarily small even if approximate inversion succeeds with non-negligible probability for an individual invocation.

\subsection{One-way state generators}

In the theoretical computer-science literature, classical-to-quantum
one-wayness is often formalized through the notion of a  
\emph{one-way state generator} (OWSG)~\cite{MorimaeYamakawa2022,CaoXue22,MorimaeYamakawaTQC2024}.
At an intuitive level, an OWSG formalizes the idea of an efficiently computable classical-to-quantum mapping together with an efficient verification procedure. 
A quantum state labeled by a secret classical value is easy to generate, whereas producing any label that passes verification remains computationally hard, even when polynomially many copies of the state are available.
The definition is deliberately phrased in a way that
interfaces smoothly with  cryptographic applications, such as
quantum digital signatures and quantum money.

An OWSG consists of three quantum algorithms
$
(\mathsf{KeyGen},\,\mathsf{StateGen},\,\mathsf{Ver}),
$
with the following roles:
\begin{itemize}
\item $\mathsf{KeyGen}$ is a (classical) key-generation algorithm that
on input of a security parameter $1^n$ outputs a classical ${\rm poly}(n)-$bit  label
$x \in \mathbb{X}$, sampled from some efficiently samplable
(not necessarily uniform) distribution.

\item $\mathsf{StateGen}$ is a QPT generator
which, on input $x$, prepares a quantum state
$\rho_x := \mathsf{StateGen}(x)$
on at most $\mathrm{poly}(n)$ qubits. This is the quantum state
actually distributed to users or adversaries.

\item $\mathsf{Ver}$ is a (possibly quantum) verification algorithm that, 
given a candidate label $x'$ and a quantum state $\rho_x$, outputs
``accept'' or ``reject''. It is required that  
legitimate pairs $(x,\rho_x)$ are always accepted, with negligible probability of rejection. 
%\[
%\Pr\bigl[\mathsf{Ver}(k,\varphi_k) = \text{accept}\bigr]
%\ge 1 - \mathrm{negl}(n),
%\]
%i.e., .
\end{itemize}

The security requirement for an OWSG is formulated as the following verification game.
Even if an adversary receives polynomially many copies of the state
$\rho_x$, generated for a secret key
$x\leftarrow\mathsf{KeyGen}(1^n)$, 
it should not be able to produce any label $x'$ that passes
verification with non-negligible probability. More precisely, for any
QPT adversary $\mathcal{A}$ that is given $c(n)$ copies
$\rho_x^{\otimes c(n)}$, where $c(n)$ is any polynomial in $n$, we
require
\[
\Pr_{x}\Bigl[
  \mathsf{Ver}\bigl(x',\rho_x\bigr) = \text{accept}
  \ \big|\ 
  x \leftarrow \mathsf{KeyGen}(1^n),\;
  x' \leftarrow \mathcal{A}\bigl(\rho_x^{\otimes c(n)},1^n\bigr)
\Bigr]
\;\le\; \mathrm{negl}(n).
\]
The probability is over the random input $x$ and the internal
randomness and measurements of $\mathcal{A}$ and $\mathsf{Ver}$. 
Operationally, one may think of an OWSG as an abstract cryptographic primitive that:
(i) internally chooses a classical label $x$, (ii) publicly emits many
copies of the associated quantum state $\rho_x$, and (iii) later
accepts or rejects proposed labels $x'$ by running the verification
procedure.

From the perspective adopted in the previous section, the OWSG security notion can be viewed as an inversion game in which the success event is defined implicitly by the verification algorithm $\mathsf{Ver}$. In contrast to Definitions~\ref{def1} and~\ref{def2}, the verifier need not implement exact-label verification or a fidelity-threshold test. Rather, the acceptance condition may be any efficiently computable relation between the candidate label and the generated quantum state.

Although closely related, Definitions~\ref{def1}–\ref{def2} and the notion of an OWSG are not formally equivalent. The first distinction concerns the number of copies made available to the adversary. Definitions~\ref{def1} and~\ref{def2} are formulated for a specified copy regime, whereas an OWSG is required to remain secure against every polynomially bounded number of copies. This stronger copy requirement reflects the intended use of OWSGs as public-key primitives, where polynomially many copies of the public quantum state may be distributed.
A second distinction concerns the notion of successful inversion. In Definitions~\ref{def1} and~\ref{def2}, success is specified through one of the inversion objectives introduced in  section \ref{sec:qowf_part2} (typically exact label recovery or fidelity-based approximate inversion). By contrast, an OWSG is defined with respect to an arbitrary efficient verification algorithm $\mathsf{Ver}$, and an adversary succeeds whenever its output is accepted by this verifier. Consequently, OWSGs should not be viewed as special cases of Definitions~\ref{def1}–\ref{def2}, nor vice versa. Rather, they represent a closely related framework in which one-wayness is defined operationally through efficient verification instead of a prescribed notion of state similarity. Nevertheless, whenever the verifier $ \mathsf{Ver}$ is implemented by a fidelity-threshold test (or any equivalent state-comparison procedure), the OWSG notion coincides with the approximate inversion framework discussed in the previous section.

In physical implementations the adversary's power necessarily increases
with the number of available copies. In the OWSG setting, information-theoretic
security cannot persist against an adversary given polynomially many copies.
Hence, it is known that statistically secure OWSGs exist in the restricted-copy
regime \cite{CavalarQuantum2025}.
In contrast, as shown in Ref.~\cite{MorimaeYamakawa2022}, an unbounded adversary given polynomially many copies can break OWSG security using shadow tomography.
Whereas full quantum state tomography aims to reconstruct an unknown (single copy) $d$-dimensional density operator and typically requires $\Theta(d^2)$ copies (up to constant accuracy), shadow tomography instead focuses on predicting how the unknown state behaves with respect to a known collection of $M>1$ two-outcome measurements $\{\hat{E}_1,\ldots,\hat{E}_M\}$ \cite{AA19}. Aaronson showed that one can estimate all $M$ acceptance probabilities $\mathrm{Tr}(\hat{E}_i \rho)$ to additive error $\epsilon$ with high confidence using only
$
c_{\rm st} = O\!\left(\epsilon^{-4}\,\log^4 M \cdot \log d\right)
$
copies of the state $\rho$~\cite{AA19}.
In the OWSG setting, one can associate a measurement $\hat{E}_x$ with
each candidate key $x\in\mathbb{X}$, corresponding to the verification
test for that key, so that $M=|\mathbb{X}|$. For exponentially large key
spaces, $\log M=\Theta(n)$, and shadow tomography therefore requires only
$\mathrm{poly}(n)$ copies to estimate the acceptance probabilities of all
candidate keys to inverse-polynomial accuracy. As shown in
Ref.~\cite{MorimaeYamakawa2022}, this can be used by a computationally
unbounded adversary to violate OWSG security with only polynomially many
copies.

In the bounded-copy information-theoretic QOWF setting introduced in
Definition~\ref{def2}, by contrast, the security notion imposes a strict
upper bound on the number of available copies. If this number lies well
below the copy complexity required by the above shadow-tomography
procedure, then this particular generic attack is not guaranteed to
succeed. This observation should not be interpreted as a lower bound on
the number of copies required for inversion, since a particular state
ensemble may admit more efficient attacks.

\subsection{Quantum money}
\label{sec:money}

While OWSGs and QOWFs focus on notions of quantum one-wayness, a closely related line of research concerns the construction of quantum money.
When discussing QOWFs the emphasis is on the hardness of inversion (preimage resistance) and on the structural relation between classical inputs and their quantum encodings. 
In contrast, the security requirements of quantum money do not rely on preimage resistance. 
Instead, they focus on unclonability, uniqueness, and unforgeability of quantum states associated with classical serial numbers.

A quantum banknote consists of a quantum state $\ket{\Psi}$,  and a classical serial number $\texttt{SN}$. There is also  an efficient verification procedure that determines whether the pair $(\ket{\Psi}, \texttt{SN})$ is valid. 
Informally, a secure scheme should guarantee that: 
(i) a banknote cannot be duplicated; 
(ii) distinct states do not verify under the same serial number; 
(iii) no adversary can mint fresh valid banknotes; and ideally 
(iv) even the mint cannot generate collisions\footnote{Requirements~(i), (ii), and~(iv) concern serial numbers corresponding to already issued banknotes. Requirement~(iii) prevents unauthorized creation of new ones.}.
These guarantees rely on the no-cloning theorem and on structural or computational uniqueness of the associated quantum states, rather than on inversion hardness. 
Thus, although quantum-money schemes are conceptually related to QOWFs, they fall outside the main QOWF framework considered in this review. 
However, for the sake of completeness, let us briefly mention some representative constructions. 

Wiesner’s original private-key scheme is based on conjugate coding discussed in section \ref{sec:const1}. That is, each banknote encodes classical data in randomly chosen bases, while a trusted authority retains the secret description for verification. 
Although information-theoretically secure against counterfeiting, the scheme requires centralized verification and therefore does not provide public verifiability.

Subsequent work has explored publicly verifiable quantum money.
Aaronson and Christiano~\cite{aaronson2012quantum} proposed constructions in which banknotes are uniform superpositions over vectors in a hidden subspace of $\{0,1\}^n$.
Security relies on the difficulty of reconstructing or duplicating this hidden structure from limited quantum access and is established in an oracle (black-box) model, where the adversary is granted access only to idealized membership oracles rather than an explicit description of the hidden subspace.
Later approaches pursued alternative mechanisms, including obfuscation-based schemes~\cite{BenDavidSattath2023,CLLZ21} and quantum lightning constructions~\cite{Zhandry2021QuantumLightning,zhandry2020upl,coladangelo2023isogenies}, where producing two valid banknotes with the same serial number would imply solving an underlying computationally hard problem.

In summary, although quantum money employs ingredients closely related to those appearing in QOWFs and OWSGs, its security notions emphasize uniqueness, unclonability, and unforgeability rather than inversion hardness. It should therefore be viewed as a complementary but distinct research direction within quantum cryptography. The focus of the present review is on primitive notions of quantum one-wayness and their underlying assumptions, rather than on the broad range of cryptographic applications that may be built upon them. The purpose of this brief subsection is therefore to place quantum money in context and highlight its conceptual connection to quantum one-wayness, without attempting a comprehensive review of the subject.

%%%%%%%%%%%%%%%%%%%%
%\newpage
%%%%%%%%%%%%%%%%%%%%%%%%%%%%%%%%%%%%%%%%%%

\section{Quantum one-wayness and Quantum indistinguishability}
\label{sec:ind}

The indistinguishability problem for a QOWF $x\mapsto \rho_x$ can be formalized as a two-case decision game. 
Here, the adversary is given access to $c(n)$ polynomially bounded copies of a quantum state corresponding to a single draw from one of two 
possible ensembles\footnote{Typically the ensembles are defined with uniform probabilities $p_x = 1/|\mathbb{X}|$ and 
$q_y = 1/|\mathbb{Y}|$.}:
\[
\mathcal{E}_0 = \{(p_x,\rho_x^{\otimes c}) : x\in\mathbb{X}\},
\qquad
\mathcal{E}_1 = \{(q_y,\sigma_y^{\otimes c}) : y\in\mathbb{Y}\},
\]
where $\mathcal{E}_1$ is a reference (ideal) ensemble. 
%In the simplest case, $\mathcal{E}_1$ may consist of a single fixed state, for example
%the average state $\bar{\rho}=\sum_x p_x\rho_x$.
The adversary must decide which of the following is true:
\begin{itemize}
\item \emph{Case 0:} the given state has originated from $\mathcal{E}_0$,
\item \emph{Case 1:} the given state has originated from $\mathcal{E}_1$. 
\end{itemize}
To this end, the adversary may perform any quantum polynomial-time operation allowed by physics on the received systems,
including collective measurements across all~$c$ copies. 
Without loss of generality, we will assume ${\mathbb X} = \{0,1\}^n$, where $n\in{\mathbb N}$ is the security parameter. 
Unless stated otherwise, we assume that the ensembles are sampled uniformly.

The classical notion of computational indistinguishability of probability
distributions extends naturally to quantum state ensembles.
\begin{definition}[Computational indistinguishability]
\label{def3}
The ensembles $\mathcal{E}_0$ and $\mathcal{E}_1$ are said to be
\emph{computationally indistinguishable}, denoted by
$\mathcal{E}_0 \underset{\mathrm{c}}{\approx} \mathcal{E}_1$, if, for every
QPT distinguisher $\mathcal{D}$,
\begin{equation}
\Big|
\Pr_{x}\!\big[\mathcal{D}(\rho_x^{\otimes c}) = 1\big]
-
\Pr_{y}\!\big[\mathcal{D}(\sigma_y^{\otimes c}) = 1\big]
\Big|
\le
\mathrm{negl}(n).
\label{Comp_Ind:Qdef}
\end{equation}
\end{definition}

A QPT distinguisher is a CPTP map followed by a two-outcome POVM, where outcome “1” denotes acceptance of the hypothesis that the state was drawn from $\mathcal{E}_1$, and outcome “0” denotes acceptance of the hypothesis that the state was drawn from $\mathcal{E}_0$.
 The probabilities are over the random choices of~$x$ and~$y$, as well as the internal randomness
and measurements of~$\mathcal{D}$. Let us take a closer look at these 
probabilities. Consider for instance  
$\Pr_{x}\!\big[\mathcal{D}(\rho_x^{\otimes c}) = 1\big]$, and let $\hat{M}_1$ be 
the POVM measurement operator associated with the result “1” . 
Then the probability for $\mathcal{D} $ to return “1”  given the state 
$\rho_x^{\otimes c}$ is $\Pr\!\big[\mathcal{D}(\rho_x^{\otimes c}) = 1\big]= {\rm Tr}(\hat{M}_1 \rho_x^{\otimes c})$. Taking into account that $x$ is chosen 
at random, we have for the average probability
\[
\Pr_x\!\big[\mathcal{D}(\rho_x^{\otimes c}) =1\big]= 
 \sum_{x\in\mathbb X} p_x
{\rm Tr}(\hat{M}_1 \rho_x^{\otimes c}) =  
{\rm Tr}(\hat{M}_1 \bar{\rho}^{(c)}) = 
\Pr\!\big[\mathcal{D}(\bar{\rho}^{(c)}) =1\big].
\]  
where $\bar{\rho}^{(c)} = \sum_x p_x\rho_x^{\otimes c}$. 
The same arguments hold for the probability $\Pr_{y}\!\big[\mathcal{D}(\sigma_y^{\otimes c}) = 1\big]$ and thus 
the definition (\ref{Comp_Ind:Qdef}) reduces to 
\bea
\Big|
\Pr\!\big[\mathcal{D}(\bar{\rho}^{(c)}) = 1\big]
-\Pr\!\big[\mathcal{D}(\bar{\sigma}^{(c)}) = 1\big]
\Big| \leq \mathrm{negl}(n),
\label{Comp_Ind_Av:Qdef}
\eea
with $\bar{\sigma}^{(c)}=\sum_y q_y\sigma_y^{\otimes c}$. The quantity on the left-hand side 
of inequalities (\ref{Comp_Ind:Qdef}) and (\ref{Comp_Ind_Av:Qdef}) is usually referred to as the “advantage” 
of the distinguisher,   
and indistinguishability means that this advantage is negligible in the security parameter $n$.
Moreover, we emphasize again that computational indistinguishability is defined 
with respect to all QPT distinguishers.

If one considers the maximal advantage over all physically allowed 
distinguishers (including those with unbounded computational 
power), this reduces to a maximum over all two-output (accept/reject) POVMs, and thus (see also appendix \ref{appendix7}) \cite{NielsenChuang2000} 
\bea
\max_{\rm POVM}\Big|
{\rm Tr}(\hat{M}_1 \bar{\rho}^{(c)}) 
-{\rm Tr}(\hat{M}_1 \bar{\sigma}^{(c)}) \Big| =\frac{1}2\left\|\,
\bar{\rho}^{(c)}
-
\bar{\sigma}^{(c)}
\right\|_1.
\label{IT_Ind_Av:Qdef}
\eea 
Hence, if  the two average states are close in  trace distance i.e., 
\[
\frac{1}{2}\left\|\,
\bar{\rho}^{(c)}
-
\bar{\sigma}^{(c)}
\right\|_1
\;\leq\; \mathrm{negl}(n).
\] 
there is no quantum measurement that can distinguish from which ensemble 
the state was drawn. In other words, the two ensembles are \emph{information-theoretically} (or \emph{statistically}) 
\emph{indistinguishable}, and we write $\mathcal{E}_0 \underset{\mathrm{inf}}{\approx} \mathcal{E}_1$. 
In particular, if $\bar{\rho}^{(c)}= \bar{\sigma}^{(c)}$ exactly, then information-theoretic indistinguishability holds with advantage~$0$.

Typically, when discussing computational indistinguishability in the classical case, the reference distribution 
 is uniform over bitstrings of a certain length. 
In the quantum setting, however, the maximally mixed state is not always an appropriate reference, since the possible states 
$\{\rho_x: x\in\mathbb {X}\}$ may for instance occupy a  subspace of the entire Hilbert space, or they may exhibit specific symmetries.
So, a meaningful reference must respect the same support and symmetry structure as $\mathcal{E}_0$; otherwise distinguishing may become trivial for artificial reasons.
As will be seen below, in many cases, setting $\sigma_y^{\otimes c}$ equal to the average state $\bar{\rho}^{(c)}$ provides a 
natural reference, analogous to the uniform distribution in the classical setting.

Let us  have a look at some of the above mentioned QOWFs, from the indistinguishability viewpoint. In construction 3, 
for the ensemble ${\cal E}_0$ we have $p_x = 1/|{\mathbb X}|$ 
and $\rho_x^{(c)} = (\ket{\psi_x}\bra{\psi_x})^{\otimes c}$ with  $\ket{\psi_x(\theta_n)}$ given by Eq. (\ref{1s}).
From the adversary's point of view, who does not have access to the uniformly random label $x$ of the given state, the  state is $\bar{\rho}^{(c)}  = |{\mathbb X}|^{-1} \sum_x  \rho_x^{(c)} $, which is equal to a Binomial mixture of the Dicke states 
$\{D_k^{c}:0\leq k\leq c\}$  (see appendix \ref{appendix4})
\begin{equation}
\bar{\rho}^{(c)} \;=\;
2^{-c}\sum_{k=0}^c \binom{c}{k}\ket{D_k^{(c)}}\!\bra{D_k^{(c)}},
\label{eq:qubit_bar-copies}
\end{equation}
which is independent of the sampled label $x$.
Consequently, the ensemble
$\mathcal E_0$
is information-theoretically indistinguishable from the reference ensemble  ${\mathcal E_1} = \{1/|{\mathbb X}|, \bar{\rho}^{(c)} \}$, 
where the challenger samples a uniformly random label
$x\in\mathbb X$, but always returns the same averaged state  $\bar\rho^{(c)}$.
This is because,  both ensembles produce exactly the same average density operator. 

For a single copy ($c{=}1$) we have $\bar{\rho}^{(1)}=\mathbb I/2$ (maximally mixed), and in this case the ensemble is
information-theoretically indistinguishable from uniform noise.
However, for $c\ge2$, the challenger still draws a \emph{pure} symmetric state $\rho_x^{\otimes c}$,
but the averaged state seen by an uninformed adversary is the binomial mixture~\eqref{eq:qubit_bar-copies}.
Any distinguisher able to detect non-symmetric components of a candidate state (e.g., by projecting onto the symmetric subspace) would trivially distinguish the QOWF output from the fully maximally mixed state (see appendix \ref{appendix4}), even though such a test reveals no information  about the hidden label $x$. This illustrates that the correct indistinguishability reference for $c\ge2$ must respect the same permutation symmetry as the QOWF outputs.

As discussed above, in the case of qubits the inversion of the mapping
$x \mapsto \rho_x^{\otimes c}$ is expected to be information-theoretically
hard whenever
$n \gg \log_2(c+1)$.
Indeed, by the Holevo bound, an adversary receiving only $c$ copies of the
state can extract at most $\log_2(c+1)$ bits of information, whereas $n$
bits are required to identify the hidden label $x$.
It is important to emphasize, however, that inversion and
indistinguishability are fundamentally different tasks.
Any adversary that successfully recovers the hidden label $x$ can certainly
determine from which ensemble the state was drawn, but the converse is not
true: an adversary may distinguish two ensembles without recovering $x$.
Moreover, the two tasks are governed by different physical mechanisms and
need not exhibit the same behavior as system parameters change.
For example, even in parameter regimes where inversion becomes possible,
distinguishing the ensemble $\mathcal E_0$ from a suitably chosen reference
ensemble may still be impossible.

The indistinguishability statement discussed here concerns only the
ensemble-identification problem.
The fact that the averaged state
$\bar{\rho}^{(c)}$
is independent of the sampled label $x$
does not imply that the individual states
$\{\rho_x^{\otimes c}\}_{x\in\mathbb X}$
cannot be distinguished or that the hidden label cannot be recovered.
Rather, it implies only that when the challenger replaces
$\mathcal E_0$
by the reference ensemble
$\mathcal E_1$,
whose every sample is the averaged state
$\bar{\rho}^{(c)}$,
no quantum measurement can distinguish between the two ensembles.
Recovering the hidden label $x$ is therefore a separate task whose
difficulty depends on the distinguishability properties of the individual
states rather than solely on their average.

Similar arguments hold for construction 4. Again, the indistinguishability argument is entirely at the level of the averaged state.   
For weak symmetric coherent states, one can use the approximation (\ref{eq:coh_qubit}), thereby obtaining that 
from the adversary's point of view, the one-copy state is diagonal in the Fock basis, while for $c$ copies, one may work 
 as in appendix \ref{appendix4} obtaining 
\begin{equation}
\bar{\rho}^{(c)} \;=\; 
(1+\mu)^{-c}\sum_{k=0}^c \mu^k \binom{c}{k}\ket{D_k^{(c)}}\!\bra{D_k^{(c)}}.
\label{eq:bar-copies-coh}
\end{equation}
In either case, the averaged state is independent of the sampled label.
Consequently, the natural reference ensemble consists of repeated copies
of the averaged state $\bar{\rho}^{(c)}$.
Since both ensembles have the same average density operator,
they are information-theoretically indistinguishable. We emphasize again that 
this indistinguishability statement concerns the ensemble-identification
problem and should not be confused with recovery of the hidden label.

The indistinguishability notion introduced above plays a central role in several modern quantum cryptographic primitives. 
In particular, pseudorandom quantum states and pseudorandom state generators are defined through computational 
indistinguishability from suitable reference ensembles. We therefore turn next to these constructions.

%%%%%%%%%%%
%\newpage
%%%%%%%%%%%

\section{Efficiently Indistinguishable Pairs of Quantum States}
\label{sec:efi}

Brakerski, Canetti, and Qian generalized Goldreich’s notion of efficiently 
indistinguishable distributions \cite{Goldreich1990CI} to the quantum setting by 
introducing {\em efficiently indistinguishable} (EFI) pairs of quantum states \cite{BCQ-ITCS2023}. 
An EFI pair consists of two efficiently preparable mixed quantum states $(\rho_0,\rho_1)$ 
satisfying two complementary properties. First, no QPT distinguisher, even when 
given $c(n)$ polynomially many copies $\rho_b^{\otimes c(n)}$ for a hidden bit $b\in\{0,1\}$, 
can distinguish $\rho_0$ from $\rho_1$ with more than negligible advantage. 
Second, the states are statistically distinguishable, i.e., their trace distance is non-negligible, so an 
unbounded adversary can distinguish them with non-negligible advantage.
 Thus, EFI pairs exhibit a 
separation between efficient and unbounded distinguishability. In particular, the two states may be 
information-theoretically far apart, for example having constant (or even close to maximal) trace distance, while remaining computationally indistinguishable. Equivalently, EFI indistinguishability does not imply information-theoretic closeness: two states forming an EFI 
pair may admit an optimal Helstrom measurement with large distinguishing advantage, provided that such a measurement 
is not efficiently implementable.

In the context of EFI pairs, the indistinguishability is defined with respect to a single hidden bit $b$, rather than a distribution over many labels, distinguishing EFI pairs from general ensemble indistinguishability discussed in the previous section.
EFI pairs capture a minimal form of computational hiding for quantum states.
If one cannot hide even a single bit against QPT distinguishers, then standard
computational primitives such as computationally hiding commitments (and hence
many higher-level protocols built from them) cannot exist. In this sense, EFI
pairs play a foundational role analogous to that of computationally
indistinguishable distributions in classical cryptography.

In their work, Brakerski \emph{et al.} relate EFI pairs to primitives such as
commitments and oblivious transfer via reductions and equivalences.
They do not, however, propose a concrete stand-alone family of EFI states, but rather EFI pairs arise implicitly within their cryptographic constructions.

Well before the work of Brakerski {\em et al.}, 
Kawachi {\em et al.} introduced a computational problem pertaining to two mixed states~\cite{KKNY2013}.
Let $n\in\mathbb{N}$ be the security parameter and let ${\mathbb S}_n$ denote the symmetric group on $n$ elements.
Let also $\mathbb{C}^{n!}$ denote the Hilbert space spanned by the computational basis 
$\{\ket{\sigma}:\sigma\in \mathbb{S}_n\}$,
where each basis state is labeled by a permutation of $n$ elements. 
Since there are $n!$ orthogonal basis states, a register (classical or quantum) must be able to index all of $n!$ possibilities, which 
requires $\lceil \log_2(n!)\rceil $ bits or qubits. 
Assuming that $n$ is even, define the subset of fixed-point-free involutions
${\mathbb K}_n = \{\kappa \in \mathbb{S}_n : \kappa^2=\mathrm{id}\ \text{and}\ \forall i\in\{1,\dots,n\},\ \kappa(i)\neq i\}$.
Each $\kappa \in {\mathbb K}_n$ can be written as a product of $n/2$ disjoint transpositions; i.e., for distinct $i_j \in \{1,\dots,n\}$ we have
$\kappa = (i_1,i_2)(i_3,i_4)\ldots (i_{n-1},i_n)$, where the transposition $(i_j,i_k)$ swaps the elements $i_j,i_k\in\{1,\dots,n\}$.

Let $x=(\kappa,b)$, with domain ${\mathbb X}={\mathbb K}_n\times\{0,1\}$, 
and consider the mapping $x \mapsto \rho_{x}$, where
\begin{equation}
  \rho_{\kappa,b} \;=\; \frac{1}{2\,n!}\sum_{\sigma\in \mathbb{S}_n} \bigl(\ket{\sigma}+(-1)^b\ket{\sigma\kappa}\bigr)\!\bigl(\bra{\sigma}+(-1)^b\bra{\sigma\kappa}\bigr).
  \label{eq:kkny-states}
\end{equation}
The basis vector $\ket{\sigma\kappa}$ is labeled by the composed permutation $\sigma\kappa$.  
%(see appendix \ref{appendix6} for an explicit example).

The state $\rho_{\kappa,b}$ can be efficiently prepared in $\mathrm{poly}(n)$ time using controlled-$\kappa$ operations. Kawachi {\em et al.} considered the following computational problem. Given ${\rm poly}(n)$ many copies  of a state 
$\rho_{\kappa,b}$, for an unknown  
$\kappa\in{\mathbb K}_n$, 
 decide whether the state is $\rho_{\kappa,0}$ or $\rho_{\kappa,1}$.
 They established the following properties: 
 (i) we can efficiently distinguish between $\rho_{\kappa,0}$ and $\rho_{\kappa,1}$ if $\kappa \in{\mathbb K}_n$ is known (trapdoor);
(ii) the average-case hardness of the problem 
over uniformly random $\kappa \in {\mathbb K}_n$ 
coincides with its worst-case hardness;
and (iii) the problem in the worst case is computationally at least as 
hard  as the Graph Automorphism problem, 
for which it is widely believed that there no QPT algorithm exists.
The one-wayness here is \emph{computational}, not information-theoretic, in the sense that even with $\mathrm{poly}(n)$ copies, no efficient 
(poly-time) distinguisher should exist, because its existence would imply 
a solution to the Graph Automorphism problem. 
%%%%%%
%%%In addition 
%%%they prove that for fewer copies (specifically $o(n\log n)$ copies), even an unbounded adversary cannot distinguish between the two ensembles. 

These properties can be exploited in cryptographic protocols such as quantum public-key encryption. The permutation $\kappa$ plays the role of the 
private key, and it is chosen at random from a uniform distribution over 
${\mathbb K}_n$. A user can prepare and publish polynomially many copies
of the \emph{public-key state} $\rho_{\kappa,0}$. 
So, if Alice wants to send Bob a one-bit message $m\in\{0,1\}$, she 
requests a fresh copy of Bob’s public-key state, and sends to Bob the state 
 $\rho_{\kappa,m}$, which is the quantum ciphertext. Note that in order to 
 transform $\rho_{\kappa,0}$ to $\rho_{\kappa,1}$, 
 she applies the publicly known procedure, which does 
 not require knowing $\kappa$. 
Bob knows $\kappa$, and can distinguish perfectly between 
$\rho_{\kappa,0}$ and $\rho_{\kappa,1}$ to recover $m$.

The construction of Kawachi {\em et al.}~\cite{KKNY2013}
provides a concrete example of computationally indistinguishable quantum states arising in a trapdoor-based cryptographic setting. 
Although not formulated as an EFI pair, the construction illustrates the same general phenomenon underlying EFI indistinguishability: two efficiently preparable quantum states may be 
information-theoretically distinguishable while remaining computationally indistinguishable for efficient adversaries. For this reason, it serves as a useful illustrative example of ideas closely related to EFI indistinguishability, despite relying on a stronger cryptographic framework.
The construction imposes similar technological challenges as quantum fingerprint states, in the sense that many-qubit states are required. Hence, it is technologically more demanding than the single-qubit and coherent-state constructions.

EFI pairs and QOWFs capture complementary notions of quantum hardness.
EFI pairs concern indistinguishability between two states that can be prepared efficiently,
whereas QOWFs concern the difficulty of recovering a classical preimage from a
 quantum encoding.  The constructions of Kawachi \emph{et al.} and
Brakerski \emph{et al.} illustrate computational hiding based on classical
complexity assumptions, while the QOWFs considered in section {\ref{sec:qowf_part2}} exploit intrinsic
quantum limitations such as non-orthogonality and restricted accessible
information.  In general, neither notion implies the other, but both serve as
foundational primitives for quantum cryptography. 
The precise relationship between EFI pairs and OWSGs remains an active topic of investigation. 
One reason is that one-wayness alone does not, in general, imply indistinguishability \cite{KT23,MMWY24}.

%===========================================================
%\newpage
%============================================================

\section{Pseudorandom Quantum States}
\label{sec:prsg}

Many applications in quantum information require quantum states that reproduce the statistical properties of Haar-random  states while remaining efficiently preparable (see appendix \ref{appendix9}). 
Such efficiently generated approximations are  referred to as pseudorandom quantum states.  Two complementary approaches have emerged. 
Quantum t-designs provide an information-theoretic approximation to Haar randomness for bounded-copy distinguishers, whereas pseudorandom quantum state generators (PRSGs) replace 
exact statistical indistinguishability by computational indistinguishability. We begin by explaining why preparation of Haar-random states is believed to require exponentially large quantum circuits.

\subsection{Complexity of preparing Haar-random states}

Preparing a generic Haar-random $n$-qubit pure state is widely believed to require exponentially large quantum circuits. The basic intuition is straightforward. 
A pure state of $n$ qubits is specified by $2^n$ complex amplitudes. After accounting for normalization and the irrelevance of the global phase, this leaves
\[
D = 2(2^n-1)=2^{n+1}-2
\]
independent real parameters.\footnote{For $n=1$, this reduces to the familiar Bloch sphere, which is specified by two real parameters.}
Thus, the manifold of pure $n$-qubit states has exponential dimension.

By contrast, an efficient quantum circuit contains only polynomially many elementary gates chosen from a fixed finite universal gate set. 
Consequently, the number of distinct states that can be generated by polynomial-size circuits grows exponentially with the circuit size, 
but remains vastly smaller than the set of all pure quantum states. Equivalently, polynomial-size quantum circuits explore only a tiny portion of the manifold of pure states.

This intuition can be made quantitative by partitioning the manifold of pure states into cells of diameter $\varepsilon$, where the distance 
may be measured using any of the standard equivalent metrics on pure states (e.g., trace distance or Euclidean distance between normalized state 
vectors modulo global phase). Since the manifold has real dimension $D=2(2^n-1)$, 
each cell occupies a volume proportional to $\varepsilon^{D}$. Consequently, covering the entire manifold requires approximately
\[
\left(\frac{a}{\varepsilon}\right)^D
=
\left(\frac{a}{\varepsilon}\right)^{2(2^n-1)}
=
\left(\frac{a}{\varepsilon}\right)^{\Theta(2^n)}
\]
cells, where $a>0$ is a constant that depends only on the geometry of the manifold. Any family of circuits capable of approximating 
arbitrary pure states to accuracy $\varepsilon$ must therefore generate at least one representative state inside each such cell.
On the other hand, the number of distinct quantum circuits of length $L$ over a fixed finite universal gate set grows at most exponentially in $L$, 
and so does the number of distinct states they can prepare. Comparing these two growth rates immediately implies that approximating every pure state to constant precision requires
$L=\Omega(2^n)$. Thus, polynomial-size quantum circuits explore only an exponentially small subset of the manifold of pure quantum states, and it is highly 
unlikely for a Haar-random state to admit an efficient preparation circuit.

This counting argument is non-constructive, but it demonstrates that high circuit complexity is the rule rather than the exception. 
Importantly, it does not identify explicit hard instances. Rather, it shows that almost all pure states lie outside the reach of efficient preparation procedures. 
Exact, or even constant-precision, Haar randomness is therefore generically unattainable in scalable quantum architectures.

These observations motivate the search for efficiently preparable ensembles that capture the essential features of Haar-random states without requiring exact Haar sampling. 
One approach seeks to reproduce only the low-order statistical moments of the Haar measure, while another replaces exact statistical indistinguishability by computational indistinguishability. 
We first discuss the information-theoretic approach based on quantum t-designs.

\subsection{Quantum $t-$designs.} 

A quantum state $t-$design is a finite ensemble of pure states ${\mathbb H}=\{|\psi_i\rangle\}$ with $ |\psi_i\rangle\in(\mathbb{C}^2)^{\otimes n} $,
 that reproduces the statistical properties of uniformly random (Haar-distributed) $n-$qubit pure states up to the first $t$ moments, 
meaning that averages of all   polynomials of degree at most $t$ in the state amplitudes are identical to those obtained 
from the Haar measure \cite{AmbainisEmerson2007,DankertCleveEmersonLivine2009,BrandaoHarrowHorodecki2016,CleveLeungLiuWang2016,NakataHircheMorganWinter2017b}. Formally, the discrete average satisfies
\[
\frac{1}{|{\mathbb H}|}\sum_i (|\psi_i\rangle\langle\psi_i|)^{\otimes t}
=
\int_{\mathrm{Haar}} (|\varphi \rangle\langle\varphi|)^{\otimes t} d\varphi,
\]
where the integral is taken over the Haar measure on the unit sphere of the n-qubit Hilbert space $({\mathbb C}^2)^{\otimes n}$.

Consequently, any (possibly computationally unbounded) protocol acting jointly on at most $t$ copies cannot distinguish a quantum $t$-design from the Haar ensemble. 
One may also define an $\epsilon$-approximate $t$-design by requiring the trace-norm distance between the corresponding $t$-th moment operators to be at most $\epsilon$ \cite{ApprDesigns23}. 
Intuitively, $t$-designs provide a resource-efficient way to generate ``pseudorandom'' quantum states without sampling from a continuous distribution, which would be impractical experimentally. 
In practice, low-order designs (especially $t=1$ and $t=2$) are readily realizable on current quantum hardware using Pauli or Clifford circuits and are routinely employed in a variety of quantum-information-processing protocols, 
whereas higher-order designs require deeper, noisier circuits and are primarily of theoretical interest.

\subsection{Pseudorandom quantum state generators}

Pseudorandom quantum state generators (PRSGs) were introduced by Ji, Liu, and Song~\cite{JiLiuSong2018} as quantum analogues of classical PRGs. 
A PRSG  is an efficient classical-to-quantum map $ x \;\mapsto\; \ket{\psi_x}$, 
where $x\in\{0,1\}^n$ is a classical seed and $\ket{\psi_x}$ is an efficiently preparable $q(n)$-qubit quantum state, where the output length $q(n)$ is polynomial in the seed length $n$. 
The defining security requirement of a PRSG is computational pseudorandomness. 
Specifically, for every polynomial number of copies $c(n)$, the ensembles
\[
\mathcal{E}_{\mathrm{PRS}}^{(c)}
=\{(|\psi_x\rangle^{\otimes c}) : x \leftarrow \{0,1\}^n\}
\quad\text{and}\quad
\mathcal{E}_{\mathrm{Haar}}^{(c)}
=\{(|\varphi\rangle^{\otimes c}) : |\varphi\rangle \leftarrow \mathrm{Haar}\},
\]
where $|\phi\rangle \leftarrow \mathrm{Haar}$ 
denotes a pure state sampled from the unitarily invariant Haar measure, 
are computationally indistinguishable in the sense of
Definition~\ref{def3}.
Thus, although the states $\ket{\psi_x}$ are produced by short, explicitly known circuits, 
\emph{no QPT algorithm can distinguish them from Haar-random quantum states}, 
even when given many copies. 
%This requirement is substantially stronger than one-wayness, because any ability to extract even partial structure from $\ket{\psi_x}^{\otimes c}$ would already lead to a distinguishing attack.

Several explicit constructions of PRSGs are known  \cite{JiLiuSong2018,BrakerskiShmueli2019,BrakerskiShmueli2020,MorimaeYamakawa2022}, 
and they rely on classical post-quantum hardness assumptions, such as the existence of quantum-secure pseudorandom functions. It remains an open question whether PRSGs can instead be based on 
intrinsically quantum complexity assumptions or physically motivated conjectures, rather than on classical post-quantum cryptographic primitives. 
For sufficiently large Hilbert space dimensions, bipartite pseudorandom quantum states inherit (up to computational indistinguishability) the near-maximal entanglement property of typical Haar-random pure states \cite{JiLiuSong2018}. 
Pseudorandom density matrices (PRDMs), introduced by Bansal et al.~\cite{PRXQuantum.6.020322}, 
extend pseudorandomness to mixed states. 
They constitute a parallel generalization of PRSGs to the mixed-state setting, rather than a strictly stronger primitive.

PRSGs and EFI pairs capture related but distinct notions of computational
indistinguishability. A PRSG compares an efficiently generated state ensemble
with the Haar ensemble, whereas an EFI pair consists of two efficiently
preparable state ensembles that are computationally indistinguishable while
remaining statistically far apart. Consequently, the two notions should not be regarded as equivalent without additional assumptions. 
To illustrate this distinction, consider the average output state of a PRSG,
\[
\rho_{\mathrm{PRS}}
=
\frac{1}{2^n}
\sum_{x\in\{0,1\}^n}
|\psi_x\rangle\langle\psi_x|,
\]
and the corresponding single-copy Haar average,
\[
\rho_{\mathrm{Haar}}
=
\int
|\varphi\rangle\langle\varphi|\,d\varphi
=
\frac{\mathbb I_d}{d},
\]
where $d=2^{q(n)}$ is the dimension of the Hilbert space.
The pseudorandomness property implies computational indistinguishability
between the generated and Haar ensembles, even when polynomially many copies
are available. However, this alone is insufficient to obtain an EFI pair,
because EFI pairs additionally require the two state ensembles to be
statistically distinguishable. 
Indeed, the average PRSG state may coincide with, or be arbitrarily close to, the Haar average $\mathbb I_d/d$. 
For example, equality holds whenever the generated ensemble forms an exact state 1-design, 
while an approximate state 1-design yields an average state that is correspondingly close to $\mathbb I_d/d$.
 Consequently, a PRSG does not automatically induce an EFI pair
through the two averaged states above. Establishing such a relation therefore
requires additional assumptions, an explicit construction, or an independent
lower bound on their statistical distance.

The relationship between quantum $t$-designs and PRSGs closely parallels the distinction between bounded-copy information-theoretic QOWFs and OWSGs discussed in section~\ref{sec:qowf_part2}. In the framework of PRSGs, the adversary may receive polynomially many copies of the generated state, but its computational power is restricted to quantum polynomial time. By contrast, quantum $t$-designs place no computational restriction on the distinguisher, but limit it to collective operations on at most $t$ copies of the state. Thus, the two notions approximate Haar randomness in complementary ways: PRSGs rely on computational indistinguishability, whereas quantum $t$-designs provide information-theoretic indistinguishability for adversaries restricted to at most $t$ copies.
These two approaches become closely related when both resources are simultaneously bounded. Kretschmer~\cite{Kretschmer2021TQC} showed that every efficient approximate $t$-design is also a $t$-copy PRSG. Moreover, Cavalar {\em et al.}~\cite{CavalarQuantum2025} proved that any efficient approximate $t$-design mapping $n$-bit seeds to $\omega(\log n)$ qubits also constitutes a $(t-1)$-copy OWSG.

Cavalar {\em et al.}~\cite{CavalarQuantum2025} analyzed in detail the
relationship between PRSGs and OWSGs. They showed that, under suitable
conditions on the output length, PRSGs satisfy the requirements of an
OWSG.
At the same time, they established fundamental limits on
information-theoretic one-wayness.
In particular, for every
$c(n)=o(n/\log n)$ there exists a $c(n)$-copy
information-theoretically secure OWSG.
By contrast, security against $\Omega(n)$ copies cannot be achieved
information-theoretically and therefore requires computational
assumptions.
Moreover, they proved that the existence of OWSGs would imply the
complexity-theoretic separation $\mathtt{BQP}\neq\mathtt{PP}$.
Here $\mathtt{BQP}$ denotes the class of problems efficiently solvable by
quantum computers, while $\mathtt{PP}$ is a much larger probabilistic
class\footnote{Here $\mathtt{PP}$ (Probabilistic Polynomial time)
denotes the class of decision problems solvable by a classical
probabilistic polynomial-time algorithm that outputs the correct answer
with probability strictly greater than $1/2$ on every input.}
that is known to contain $\mathtt{BQP}$ and is widely believed to be
strictly more powerful.
Proving $\mathtt{BQP}\neq\mathtt{PP}$ would therefore constitute a major
complexity-theoretic breakthrough\footnote{Since it is known that
$\mathtt{BQP}\subseteq\mathtt{PP}$~\cite{AdlemanDeMarraisHuang1997},
the separation $\mathtt{BQP}\neq\mathtt{PP}$ would imply the strict
containment $\mathtt{BQP}\subsetneq\mathtt{PP}$.}.
Consequently, this implication provides evidence that unconditional
(i.e., assumption-free) constructions of OWSGs are unlikely to exist.
Taken together, these results suggest that few-copy one-wayness can
arise purely from information-theoretic limitations, whereas achieving
OWSG security against polynomially many copies requires computational
assumptions.

PRSGs, OWSGs, and computational QOWFs all map classical inputs to efficiently generated quantum states, but they differ in the security guarantees they provide. A computational QOWF, as introduced in Definition~\ref{def1}, requires inversion hardness with respect to a specified copy regime. By contrast, an OWSG strengthens this notion by requiring one-wayness to hold even when the adversary is given any polynomially bounded number of copies of the generated state, together with an efficient verification procedure. PRSGs impose a qualitatively stronger computational requirement: the generated states must be computationally indistinguishable from Haar-random states. The results of Cavalar \emph{et al.} establish that, under suitable conditions on the output length, every PRSG is also an OWSG. Accordingly, under compatible choices of the inversion objective and verification procedure, one obtains the conceptual hierarchy
\[
\text{PRSG}
\;\Longrightarrow\;
\text{OWSG}
\;\Longrightarrow\;
\text{computational QOWF},
\]
whereas the converse implications do not hold in general. 
This conceptual hierarchy should not be confused with the bounded-copy
information-theoretic QOWFs of Definition~\ref{def2}, several examples of
which were discussed in Sec.~\ref{sec:qowf_part2}. Those constructions derive
their security from fundamental limits on quantum state discrimination and
accessible information rather than from computational hardness. Consequently,
they often generate highly structured state families that may be efficiently
distinguishable from Haar-random states, even though recovering the classical
input remains information-theoretically hard in the relevant bounded-copy
regime.

While PRSGs, PRDMs, and EFI pairs are primarily theoretical primitives,
they provide complementary computational notions of quantum
pseudorandomness and play a foundational role in quantum cryptography and
quantum complexity theory. PRSGs aim to emulate Haar-random quantum
states, whereas EFI pairs capture computational indistinguishability
between efficiently preparable state ensembles that remain statistically
distinguishable. By contrast, the information-theoretic QOWF
constructions discussed earlier derive their security from fundamental
limitations on quantum information processing rather than from
computational hardness, making them particularly relevant for
near-term photonic implementations.

%===========================================================
%\newpage
%============================================================

\section{Other Constructions of Quantum One-Way Functions}
\label{sec:other}

While the canonical formulation of a QOWF is a
classical-to-quantum one-way mapping $k \mapsto \rho_k$,
one-wayness has also been investigated in settings involving quantum inputs.
We briefly review two such directions: quantum-to-classical and quantum-to-quantum OWFs.

\subsection{Quantum-to-classical maps}
Behera and Paul \cite{BeheraPaul2018} presented a quantum-to-classical OWF $f_{qc}$, which takes as input an 
$n$-qubit state chosen at random from the so-called GCH$_n$ set of $n$-qubit states. A GCH state is a tensor product state, and involves blocks of qubits 
in a GHZ state from the set $\{(\ket{i_1,i_2,\ldots,i_n}+\ket{\bar{i}_1,\bar{i}_2,\ldots,\bar{i}_n})/\sqrt{2}:\, i_j\in\{0,1\}, \bar{i}_j = i_j\oplus 1\}$, and blocks of qubits in product states of $\{|0\rangle,|1\rangle, |+\rangle,|-\rangle\}$. 
The goal of the construction is to map a restricted family of quantum states onto deterministic classical bit strings while making inversion difficult.

For the evaluation of $f_{qc}$ one applies layers of CNOT gates on
``compatible'' qubit pairs. Here, compatibility refers to pairs of qubits
for which the action of the CNOT gate preserves membership of the resulting
state in the GCH$_n$ family. Thus, after each layer of CNOT operations,
the remaining unmeasured state is still a valid GCH$_n$ state and can be
processed further by the protocol.
Projective measurements are then performed on each qubit which is either in  $\{|0\rangle,|1\rangle\}$ or in 
$\{ |+\rangle,|-\rangle\}$ basis, thereby obtaining a  bit string deterministically determined by the input state and chosen measurement bases. 
At the end of each layer unmeasured qubits may remain entangled, and they 
enter the next layer of CNOT gates. The hierarchy of layers ends when no compatible qubits remain, and results in an $\ell$-bit string, where $\ell\leq n$ equals the total number of qubits measured across all layers.

For the design of the Behera-Paul QOWF, one has to use a classical computer, in order to 
find pairs of compatible qubits, as well as the correct measurement bases at the end of each layer, so that the outcomes are deterministic. The design 
and the performance of $f_{qc}$ for a given input state 
is a computationally easy task. 
On the contrary, inverting the map (i.e., finding a valid input state for a 
given $\ell$-bit string) appears to be a hard task, because an adversary would need to search through a number of possible input states that scales exponentially in $n$. However, no reduction to a well-studied hard problem is provided, so the security claim remains heuristic.  Moreover, since $f_{qc}$ produces an $\ell$-bit output with $\ell \leq n$, collisions are guaranteed to exist, but it remains unclear whether they can be found efficiently. This weakens the cryptographic interpretation of $f_{qc}$ compared to established one-way or hash functions, where both preimage and collision resistance are required.

The scheme of Behera and Paul represents an early attempt to formalize a 
quantum-to-classical OWF. It is conceptually interesting, in that it 
combines structured families of quantum states (GCH states), carefully designed 
measurements, and classical preprocessing to obtain deterministic classical outputs, 
while invoking no-cloning and other quantum principles to argue about 
non-invertibility. However, the construction should be viewed primarily as a 
conceptual stepping stone rather than a cryptographically solid primitive. 
Its one-wayness is argued heuristically rather than reduced to a standard hard 
problem, collision resistance is not analyzed despite inevitable collisions due 
to output compression, and its proposed application to quantum money ultimately 
relies on classical digital signatures for security. In addition, major 
practical challenges such as state preparation, transmission, storage, and 
noise resilience remain unaddressed. As such, the work highlights both the 
potential and the limitations of current approaches to quantum-to-classical 
one-way functions, and underscores the need for more rigorous and 
implementable constructions.

\subsection{Quantum-to-quantum maps}

Shang et al.~\cite{shang2020full} introduced the notion of a \emph{full
QOWF}, in which both the input and output are
quantum states. Their construction relies on a combination of the  
quantum-to-classical OWF of Behera and Paul  with single-qubit rotations, 
and is efficiently computable, while inversion is argued to be computationally
hard for QPT adversaries within their security model.
The authors apply this primitive
to a quantum identity authentication protocol based on a
challenge--response mechanism.

A central component of the verification procedure is the SWAP test,
which compares the prover’s response with a stored verification state.
Because the SWAP test has one-sided error, the protocol repeats the
verification across multiple independent challenges to suppress false
acceptance. This repetition requires multiple copies of the same
verification state and is explicitly built into the security model,
with the overall forging probability decreasing exponentially in the
number of rounds.

While conceptually appealing, the construction relies on idealized
assumptions such as noiseless quantum channels. Extensions to noisy
settings require either quantum error correction or acceptance
thresholds over repeated tests, again presupposing access to multiple
copies of the verification state. Releasing many copies of a single
quantum state raises concerns about cumulative information leakage,
highlighting the need for copy-sensitive security analyses or designs
that avoid exposing verification states altogether. Moreover, the
scope of applications is limited to identity authentication, and the
practical feasibility on near-term quantum hardware remains unclear.

\subsection{Boson sampling}
\label{sec:bs-owf}

Boson sampling is a computational task widely regarded as a representative paradigm for demonstrating quantum advantage. It pertains to sending \(N\) indistinguishable bosons into an \(M\)-mode linear interferometer (\(M>N\)), implementing a Haar-random unitary matrix \(U\), and sampling from the resulting output distribution. Since its formalization by Aaronson and Arkhipov~\cite{AA2013BS}, substantial experimental effort has been devoted to scaling boson sampling and its variants~\cite{LBR2017QS,WangPan2019PhotonicReview}.

Coarse-grained boson sampling has been also exploited in the design of cryptographic OWFs \cite{NikBro2016,Nikolopoulos2019,Wang-etalBS23}. In these constructions, a classical input (e.g., specifying the boson configuration at the input of the interferometer) is mapped, via a boson-sampling experiment, to a coarse-grained classical output derived from the sampled boson distribution. 
Coarse-graining is essential to obtain a reproducible output, with sample sizes that scale polynomially in $(M,N)$, 
but necessarily introduces collisions.

Forward evaluation of the function corresponds to executing the boson-sampling process and recording the coarse-grained outcome. For small system sizes ($N\lesssim 40$), limited instances of the forward map can be simulated classically, whereas inverting the function amounts to finding a valid preimage consistent with a given output \cite{NikBro2016,Nikolopoulos2019,Wang-etalBS23}. No efficient inversion algorithms are currently known, either classical or quantum. 
The best known generic inversion strategies rely on exhaustive search over
possible inputs, with computational cost exponential in the number of bosons.
Collision-finding attacks are likewise exponential, with birthday-type
strategies reducing the exponent only quadratically. No polynomial-time
classical or quantum inversion algorithms are known.

The security of boson-sampling-based OWFs rests on the conjectured average-case hardness of computing or approximating boson-sampling amplitudes (permanents of complex matrices). Importantly, practical security depends on the robustness of the construction to experimental imperfections such as loss, partial distinguishability, and detector noise, which may alter the induced distribution and affect both correctness and resistance to inversion. As a result, while boson-sampling-based OWFs are conceptually appealing and well aligned with near-term photonic platforms, their cryptographic strength remains conditional and sensitive to implementation details.

A related line of research, distinct from boson-sampling-based OWFs, exploits coarse-grained boson sampling as a source of computational indistinguishability between classical probability distributions, leading to proof-of-principle cryptographic constructions including authentication and secure communication \cite{Nikolopoulos2023BS}.
This illustrates that sampling-based quantum advantage may provide cryptographic resources beyond one-wayness itself, suggesting an alternative route toward quantum cryptographic primitives based on computational indistinguishability rather than inversion hardness.

%===========================================================
%\newpage
%============================================================

\section{Physical Unclonable Functions}
\label{sec:puf}

Physical unclonable functions (PUFs) constitute one of the most prominent hardware-based cryptographic primitives \cite{Pappu2002POWF,McGrath2019PUFTaxonomy,Gao2020PUF,Ruhrmair2022SecretFreeSecurity,Klau-AM25}. 
Typically they pertain to physical systems whose challenge–response behavior is determined by intrinsic, uncontrollable manufacturing variations.
Given a challenge \(C\),  a PUF produces a response \(R = \mathrm{PUF}(C)\) that is easy to evaluate
given access to the device, yet difficult to clone or emulate given only
limited physical or query access.
The security of a PUF is rooted primarily in physical complexity and
manufacturing disorder rather than in computational hardness assumptions.

Quantum PUFs (QPUFs) \cite{Arapinis2021QPUF,JuliaFarre2025QPUF} extend the concept of PUFs to the quantum domain by exploiting fundamental quantum-mechanical principles, most notably the no-cloning theorem and measurement disturbance. A QPUF is a physical quantum device that implements a challenge--response behavior which is easy to evaluate given physical access to the device, yet hard to reproduce or emulate without such access due to fundamental
quantum-mechanical limits such as the no-cloning theorem and measurement disturbance. In contrast to classical PUFs, where challenges and responses are classical strings, QPUFs typically involve quantum challenges, represented by quantum states drawn from a prescribed ensemble. Upon application of a challenge state to the device, the QPUF produces a response that may be quantum or classical, depending on the measurement scheme. Formally, a QPUF can be modeled as a quantum channel $\mathcal{E}_\theta : \rho_c \mapsto \mathcal{E}_\theta(\rho_c)$,
where $\rho_c$ denotes the quantum challenge and the parameter $\theta$ encapsulates unknown and/or uncontrollable physical parameters of the device. 

 An intermediate class between classical PUFs and QPUFs is given by quantum-readout PUFs (QR-PUFs) \cite{QR_PUF10,Goorden2014,NikDia2017}. The most prominent paradigm of QR-PUFs pertains to a random optical scattering medium. 
 Each classical challenge determines the preparation of a corresponding non-orthogonal quantum state of light, which is then scattered by the random medium, 
 yielding a new quantum state that depends on both the challenge and the internal random disorder of the medium.  
Unlike QPUFs, QR-PUFs do not require the storage or transmission of long-lived quantum states, making them significantly more practical with current technology. At the same time, they inherit stronger security guarantees than purely classical PUFs, as an adversary cannot arbitrarily increase measurement precision due to
quantum limits such as shot noise, non-orthogonality of optical states, and the no-cloning principle.

A fundamental distinction between PUFs (quantum or classical) and OWFs lies in the status of the underlying mapping. In cryptographic OWFs, including their quantum variants, the mathematical description of the function is assumed to be public. Security relies on the difficulty of inverting the mapping even when the function description and an efficient procedure for evaluating the function are publicly available. In contrast, both classical and quantum PUFs implement an implicit, device-specific mapping that is not known \emph{a priori} and is assumed to be difficult to learn or emulate given limited physical access. Consequently, PUF security collapses if the mapping can be fully characterized, whereas the security of one-way functions does not depend on hiding the function description. This structural difference places PUFs and QOWFs in fundamentally distinct security paradigms.
Although early literature referred to PUFs as physical one-way functions~\cite{Pappu2002POWF}, it is now widely recognized that PUFs do not satisfy the defining properties of cryptographic one-way functions, since their security relies on limited access and restricted learnability of the underlying physical mapping.

\begin{sidewaystable}
\centering
\caption{Comparison of the main quantum cryptographic primitives discussed in this review.}
\label{tab:primitives}
\renewcommand{\arraystretch}{1.25}
\small
\begin{tabular}{p{2.5cm}p{3cm}p{3.4cm}p{4cm}p{5.0cm}}
\hline
\textbf{Primitive} &
\textbf{Efficient map} &
\textbf{Adversarial task} &
\textbf{Security model} &
\textbf{Security basis} \\
\hline

Bounded-copy information-theoretic QOWF &
$x\mapsto\rho_x$ &
Exact-label recovery or approximate inversion &
Arbitrary quantum adversary; bounded copies &
Limited distinguishability and accessible information from a bounded number of quantum states \\[0.2cm]

OWSG &
$x\mapsto\rho_x$ &
Produce a label $x'$ accepted by $\mathsf{Ver}$ &
QPT adversary; polynomially many copies &
Computational hardness of producing an accepting label \\[0.2cm]

EFI pair &
$b\in\{0,1\}\mapsto\rho_b$ &
Distinguish two efficiently preparable states &
QPT adversary; polynomially many copies &
Computational indistinguishability despite information-theoretic distinguishability \\[0.2cm]

PRSG &
$x\mapsto|\psi_x\rangle$ &
Distinguish generated states from Haar-random states &
QPT adversary; polynomially many copies &
Computational indistinguishability from Haar-random states (typically based on post-quantum assumptions) \\[0.2cm]

Quantum $t$-design &
Efficient ensemble generation &
Approximate Haar randomness &
Unbounded adversary; at most $t$ copies &
Information-theoretic matching of the first $t$ Haar moments \\[0.2cm]

PUF / QPUF &
Challenge $\mapsto$ response &
Predict the response to a previously unseen challenge &
Limited physical and/or query access &
Device-specific physical randomness; quantum unclonability (QPUFs)\\
\hline
\end{tabular}
\end{sidewaystable}

\section{Quantum One-Wayness in the Era of Post-Quantum Cryptography}
\label{sec:qowf_pqc}

An important question is how the quantum cryptographic primitives discussed in this review relate to the rapidly developing field of PQC.
 Classical post-quantum schemes aim to provide computational security against quantum adversaries while remaining entirely classical. They do not require quantum state preparation, quantum memories, quantum processors, or quantum communication channels, and they are largely compatible with existing communication infrastructures. Consequently, any quantum cryptographic primitive that provides only computational security must ultimately be evaluated against this practical benchmark, taking into account the quantum resources needed for its implementation. 
 %This observation is particularly relevant for computational primitives such as OWSGs, EFI pairs, and PRSGs, whose known constructions are themselves based on classical post-quantum assumptions. Their added value therefore lies not in replacing PQC, but in enabling cryptographic functionalities based on quantum states that have no direct classical counterpart.

From this perspective, the main promise of QOWFs and related primitives is unlikely to lie solely in reproducing computational security through more demanding quantum technologies.  Rather, their most distinctive potential advantages appear to lie in security guarantees or functionalities that have no direct classical counterpart.  These include information-theoretic or everlasting security\footnote{Everlasting security refers to security guarantees that remain valid even if an adversary acquires unlimited computational power after the protocol has been completed, provided that the adversary was computationally bounded during the execution of the protocol.}, bounded-copy security, unclonability, and other forms of security rooted in fundamental physical principles. In this sense, the practical relevance of QOWFs is closely tied to whether the quantum nature of the states can be harnessed to achieve something beyond what is naturally available from classical post-quantum assumptions.

This distinction also clarifies the role of the number of copies made available to an adversary. In classical cryptography, public values can usually be copied without restriction. In quantum cryptography, however, copies of an unknown quantum state constitute a physical resource. Increasing the number of available copies may increase the adversary's ability to extract information, perform tomography, or implement collective measurements. For this reason, bounded-copy security is not merely a technical assumption, but a physically meaningful constraint. In any realistic implementation, the number of states distributed, stored, or exposed to adversarial access is finite, and the security analysis should explicitly account for this limitation.  This viewpoint is closely related to the bounded- and noisy-storage frameworks \cite{BoundStorage05,NoisyStorage08}, in which security arises from limitations on the adversary’s ability to store quantum information. While bounded-copy and bounded-storage assumptions are formally distinct, both reflect the broader principle that quantum information is a physical resource whose availability may be constrained by realistic technological limitations. In this sense, they provide complementary approaches to obtaining security from physical restrictions rather than solely from computational hardness assumptions.

Within this perspective, classical-to-quantum QOWFs provide a natural starting point for practical quantum one-wayness. Their security may arise from limits on state discrimination, restricted accessible information, measurement disturbance, and the impossibility of cloning unknown quantum states. Such mechanisms are most directly relevant in bounded-copy regimes, where the adversary receives only a limited number of copies of the challenge state. 
This makes QOWFs whose security derives from fundamental quantum-mechanical principles, rather than solely from computational assumptions, particularly attractive  candidates for practical quantum cryptographic applications.  From an implementation perspective, however, significant differences exist among the various constructions. The single-qubit and coherent-state schemes discussed in sections \ref{sec:const1}, \ref{sec:const3}, and \ref{sec:const4} rely on technologies that are closely related to those developed for QKD, including the preparation and transmission of non-orthogonal quantum states and their subsequent measurement. Consequently, proof-of-principle demonstrations appear feasible with present-day photonic technology. At the same time, these schemes inherit many of the practical challenges familiar from QKD, including imperfect state preparation, lossy channels, detector inefficiencies, finite statistics, and the absence of ideal single-photon sources.
The situation becomes more demanding when one considers higher-level cryptographic applications. While the generation and verification of the underlying quantum states may be feasible with current technology, applications such as digital signatures, quantum public-key cryptography, or other networked cryptographic functionalities  often require the storage, repeated use, or controlled distribution of quantum states. This naturally introduces the need for quantum memories, whose performance remains a major technological bottleneck.
Additional challenges arise for constructions based on high-dimensional quantum states. In particular, the fingerprint-state construction discussed in section \ref{sec:const2} requires the generation, transmission, and faithful manipulation of large-scale coherent quantum superpositions. Such requirements go significantly beyond those encountered in current QKD systems and remain technologically demanding with present-day quantum hardware.

OWSGs represent a stronger theoretical notion, requiring one-wayness even when the adversary is given polynomially many copies of the generated state. This is a natural and powerful requirement from a complexity-theoretic perspective, and it clarifies the boundary between information-theoretic and computational notions of quantum one-wayness. At the same time, from a physical perspective, arbitrary polynomial-copy security is a demanding requirement and may not always correspond to the most natural model for near-term implementations. OWSGs should therefore be viewed as an important theoretical strengthening of QOWFs, rather than necessarily as the most immediate model for practical realization.

EFI pairs and PRSGs play a different role. They are primarily indistinguishability-based primitives, capturing computational hiding and pseudorandomness of quantum states. EFI pairs formalize the possibility that two efficiently preparable quantum states are distinguishable in principle, but not by efficient quantum algorithms. PRSGs impose an even stronger requirement, namely that efficiently generated states be computationally indistinguishable from Haar-random states. These notions are foundational for the theoretical understanding of quantum cryptography. However, their direct physical interpretation and practical deployment are  currently less straightforward than those of classical-to-quantum QOWFs.  In particular, the requirement of pseudorandomness relative to Haar-random states is sensitive to realistic noise and imperfections, which may render otherwise well-designed implementations distinguishable. 
Recent work \cite{Haug2025pseudorandom} has further highlighted this challenge by showing that pseudorandom quantum states and pseudorandom unitaries are extremely sensitive to noise. In particular, depolarizing noise that is not exponentially suppressed already renders them efficiently distinguishable from their Haar-random counterparts. Consequently, cryptographically secure realizations require error rates that decrease to negligible levels with system size. Such results suggest that certain forms of quantum pseudorandomness may lie beyond the capabilities of noisy intermediate-scale quantum devices, reinforcing the importance of explicitly accounting for physical noise and implementation imperfections when evaluating the practical relevance of quantum cryptographic primitives. 
Thus, as one moves from QOWFs to OWSGs, EFI pairs, and PRSGs, the primitives become increasingly powerful from a complexity-theoretic viewpoint, while the requirements for physically realizing their intended security properties become correspondingly more stringent. This tension between cryptographic strength and physical realizability is likely to play an important role in determining which quantum primitives ultimately prove useful in practice.

These considerations suggest that future progress should not be judged solely by the strength of abstract cryptographic definitions, but also by the availability of realistic physical implementations. For quantum one-wayness to become practically useful, one must identify state families, verification procedures, and access models for which the claimed security advantages survive loss, noise, finite statistics, imperfect state preparation, and limited quantum storage. Understanding which quantum primitives can provide genuine advantages over PQC under such constraints remains a central open problem.

\section{Conclusions and Outlook}
\label{sec:outlook}

A central message of this work is that, while the theoretical study of quantum
cryptographic primitives has advanced significantly (clarifying the
relationships between notions such as quantum one-way functions, one-way state
generators, pseudorandom quantum states, and efficiently indistinguishable state
pairs), much less progress has been made toward constructions that are both
practical and experimentally meaningful. In contrast to classical cryptography,
where abstract ideas such as one-way functions rapidly led to widely deployable
schemes, the quantum setting faces substantially higher physical and engineering
barriers. As a result, practical quantum cryptography remains largely limited to
QKD and a few closely related tasks, while many other
foundational primitives continue to exist primarily at a theoretical level.

Our analysis suggests that some of these difficulties partly reflect
intrinsic features of quantum information rather than merely temporary
technological limitations. In quantum systems, extracting information from a
state generally disturbs or destroys it, unknown states cannot be copied, and
accurately characterizing quantum processes requires many controlled
experiments. These features restrict how often quantum outputs can be reused or
replicated and make it difficult to treat quantum cryptographic primitives as
freely reusable building blocks in the classical sense. Consequently, certain
ideas familiar from classical cryptography, such as one-way functions that can
be evaluated and reused without restriction, may not always admit direct or
physically meaningful counterparts in the quantum setting without revising the
associated security expectations.

%Many proposed quantum one-way constructions therefore rely on assumptions that
%differ qualitatively from those of their classical counterparts, such as limits
%on the number of available copies of a state or on the extent of an adversary’s
%access. While such assumptions reduce generality and limit composability, they
%reflect unavoidable physical constraints rather than shortcomings of specific
%proposals. From a practical standpoint, resolving increasingly fine theoretical
%distinctions between abstract primitives offers limited immediate practical
%benefit on its own unless accompanied by constructions that can plausibly be
%implemented on near-term quantum hardware. 

Future progress in quantum cryptography will depend on a shift in
emphasis toward simple, robust, and physically realizable primitives with
clearly stated and context-dependent security guarantees that go beyond PQC. 
Even restricted or task-specific constructions, if experimentally accessible and well understood,
may play a crucial role in extending quantum cryptography beyond key
distribution, and toward a broader and more practical quantum cryptographic
toolkit.

Looking ahead, several concrete research directions appear particularly promising from a
theoretical and experimental perspective:

\begin{itemize}
\item Exploit fundamental quantum-mechanical limits (such as restricted accessible information,
measurement disturbance, and the no-cloning principle) directly as the source of one-wayness,
rather than relying exclusively on computational assumptions. This naturally motivates
bounded-copy security models, where the number of available copies of the unknown state is treated as a
physical resource and protocols are engineered to limit state reuse.

\item Better understand what quantum cryptographic primitives can offer beyond PQC. Since PQC already provides computational security against quantum adversaries without requiring quantum communication or quantum memories, it remains an important open question which functionalities or security guarantees genuinely benefit from the use of quantum states.

\item Emphasis should be given on experimentally implementable platforms, including low-dimensional encodings,
weak coherent states of light, and prepare-and-measure architectures that avoid long-term
quantum memories and complex collective measurements.

\item Incorporate realistic imperfections (loss, decoherence, detector noise, and finite statistics) 
explicitly into security analyses, replacing idealized fidelity tests with robust threshold-based
verification procedures.

\item Explore security models based on limited or noisy quantum storage, in the design of QOWFs and other primitives. 
%Bounded- and noisy-storage frameworks \cite{BoundStorage05, NoisyStorage08} 
%derive security from physical restrictions on an adversary’s 
%ability to store quantum information reliably over time, rather than from computational assumptions. 
Given the fragility and short coherence times of current quantum memories, 
such models are both theoretically appealing and experimentally realistic, 
and may enable primitives whose security is grounded directly in hardware limitations.

\item Optimize simple and physically accessible state families (e.g., phase-encoded states, coherent states or fingerprint
states), in the framework of QOWFs and other cryptographic primitives

\item Explore further sampling-based approaches to the design of cryptographic primitives. In particular, it remains unclear 
which types of sampling problems can provide useful hardness assumptions (e.g., producing or verifying samples from complex quantum distributions), and how such guarantees should be formulated when only approximate samples are available and implementations exhibit statistical noise. 

\item Hardware-rooted primitives such as QPUFs and QR-PUFs lie outside the scope of this work, in the sense that our framework studies publicly specified, efficiently computable functions (e.g., QOWFs), whereas all types of PUFs rely on hidden, device-specific randomness and therefore operate under a fundamentally different security model. Nevertheless, such hardware-based approaches constitute an interesting and complementary direction, enabling a variety of cryptographic primitives based on physical unclonability rather than computational hardness.

\item Developing task-specific quantum cryptographic primitives beyond QKD, equipped with practical and realistically achievable security guarantees (such as bounded/noisy quantum storage assumptions or limited-copy public 
quantum states) constitutes an important open challenge.

\end{itemize}

Beyond their intrinsic theoretical interest, QOWFs and related primitives may provide a pathway toward cryptographic functionalities that extend well beyond QKD. 
As experimental quantum technologies continue to mature, understanding which of these primitives can be realized in practice may play an important role in the development of a broader quantum-cryptographic ecosystem. Answering this question will likely require a closer integration of complexity-theoretic, information-theoretic, and experimental perspectives.

More broadly, an important open question is how to balance theoretical rigor with physical
implementability. Which cryptographic functionalities can be achieved with minimal quantum
resources, and what level of security is both meaningful and experimentally attainable in realistic
quantum networks?

In closing we note that many additional open questions concern the formal power, limitations, and
interrelations of the various quantum cryptographic primitives themselves, such as QOWFs, OWSGs, 
PRSGs, and EFI pairs, including questions of reductions, separations, and complexity-
theoretic assumptions. These problems are primarily rooted in information and computer
science and are treated extensively in the theoretical literature cited throughout this review and the references therein.
In contrast, the present article deliberately adopts a physics-oriented perspective, emphasizing
questions of physical realizability, noise robustness, and experimental feasibility. Our aim is to
highlight those aspects of quantum one-wayness that can be understood, tested, and ultimately
implemented in realistic quantum systems.

\ack{The publication of the article in OA mode was financially supported by HEAL-Link.}

\funding{This research was co-funded by the European Union under the Digital Europe Program grant agreement number 101091504.}
% This section is a list of funder names and grant numbers

\section*{Conflicts of Interest}
The authors declare no conflicts of interest.

\data{All data that support the findings of this study are included within the article (and any supplementary files).}

%\vspace{0.5em}
%\noindent\textbf{Acknowledgements.} [Add funding and collaborations.]

\newpage

\appendix

\section{Asymptotic notation}
\label{appendix2}

For clarity, we briefly review the standard asymptotic notation used in the
information-theoretic statements of this work.  These notations describe the
growth of a function $f(n)$ with respect to another function $g(n)$ as the input parameter 
$n$ (typically measuring problem size) becomes large.

\paragraph{Big-$O$ notation.}
Big-$O$ provides an \emph{upper bound} on the growth of function $f(n)$, relative to another function $g(n)$. 
In particular, we write $f(n) = O(g(n))$ if 
$f(n)$ does not grow \emph{faster} than $g(n)$, up to a constant factor $c > 0$, that is 
\[
  |f(n)| \le c\,|g(n)| \qquad \text{for all } n > N.
\]
for $N \in \mathbb{N}$. 

\paragraph{Big-$\Omega$ notation.} 
Big-$\Omega$ provides a \emph{lower bound} on the growth of $f(n)$, relative to $g(n)$. 
We write $f(n) = \Omega(g(n))$ if there exist constants $c > 0$ and
$N \in \mathbb{N}$ such that
\[
  |f(n)| \ge c\,|g(n)| \qquad \text{for all } n > N.
\]
In this case, $f(n)$ grows at least as fast as $g(n)$, up to a constant factor.

\paragraph{Big-$\Theta$ notation.}
We write $f(n) = \Theta(g(n))$ when $f(n)$ and $g(n)$ grow at the same rate up to fixed
multiplicative constants.  Big-$\Theta$ therefore expresses a \emph{tight bound}:
\[
  c_1 g(n) \le f(n) \le c_2 g(n) \qquad \text{for all } n > N,
\]
for some constants $c_1, c_2 > 0$.

\paragraph{Little-$o$ notation.}
When 
\[
  \lim_{n\to\infty} \frac{f(n)}{g(n)} = 0, 
\]
the function $f(n)$ grows \emph{strictly slower} than $g(n)$. In
particular, unlike Big-$O$, little-$o$ does not allow $f(n)$ to grow at
the same asymptotic rate as $g(n)$ even up to a constant factor. In this case we write $f(n) = o(g(n))$. 

\paragraph{Little-$\omega$ notation.}
We write $f(n) = \omega(g(n))$ if
\[
  \lim_{n\to\infty} \frac{f(n)}{g(n)} = \infty.
\]
This means that $f(n)$ grows \emph{strictly faster} than $g(n)$, in the
sense that $f(n)$ eventually dominates $g(n)$ by an unbounded factor.
%
%\begin{itemize}
%  \item \textbf{Polynomial example.} Let $h(n) = 3n^2 + 5n + 2$. Then
%        \[
%          h(n) = O(n^2), \qquad h(n) = \Omega(n^2), \qquad h(n) = \Theta(n^2).
%        \]
%        The quadratic term $3n^2$ dominates the expression.
%
%  \item \textbf{Logarithmic example.} Let $f(n) = 7\log n + 10$. Since the
%        additive constant does not affect asymptotic growth,
%        \[
%          f(n) = \Theta(\log n).
%        \]
%
%  \item \textbf{Mixed linear-logarithmic example.} Let $g(n) = n\log n + 100n$.
%        The $n \log n$ term dominates the linear term, and hence
%        \[
%          g(n) = \Theta(n\log n).
%        \]
%
%  \item \textbf{Exponential versus polynomial.} Let $k(n) = 2^n + n^3$. Since
%        exponential growth eventually dwarfs polynomial growth,
%        \[
%          k(n) = \Theta(2^n).
%        \]
%\end{itemize}

These notations allow us to express scaling properties cleanly when comparing,
for example, the length of a classical string and the amount of accessible
information an adversary may extract from quantum states.

\section{Asymptotics of  $D_{\mathrm{sym}}(c,d)$}
\label{appendix3}

Consider $c$ identical $d$-level systems (qudits) with computational basis 
$\{\ket{0},\ket{1},\dots,\ket{d-1}\}$ for $\mathbb{C}^d$. 
A general basis vector of the symmetric subspace is specified by occupation numbers 
$(n_0,n_1,\dots,n_{d-1})$ with $n_i\ge 0$ and $\sum_{i=0}^{d-1} n_i = c$:
\[
\ket{n_0,\dots,n_{d-1}} 
= \frac{1}{\sqrt{\frac{c!}{n_0!n_1!\cdots n_{d-1}!}}}
\sum_{\pi\in {S}_c} \hat{P}_\pi
\big(\ket{0}^{\otimes n_0}\otimes\ket{1}^{\otimes n_1}\otimes\cdots\otimes\ket{d-1}^{\otimes n_{d-1}}\big),
\]
where $S_c$ is the symmetric group on $c$ elements, whose members
$\pi$ correspond to all possible rearrangements of the $c$ subsystems,
and $\hat{P}_\pi$ is the corresponding permutation operator acting on
$(\mathbb{C}^d)^{\otimes c}$.
Each distinct integer tuple $(n_0,\dots,n_{d-1})$ corresponds to one linearly independent 
symmetric basis vector, so the dimension $D_{\mathrm{sym}}(c,d)$ of the symmetric subspace equals the number 
of nonnegative integer solutions to $\sum_{i=0}^{d-1} n_i=c$. 
In the language of combinatorics we have $c$ indistinguishable balls and $d$ distinct bins, and thus 
\[
D_{\mathrm{sym}}(c,d)=\binom{c+d-1}{d-1}.
\]

We are interested in the asymptotics of $D_{\mathrm{sym}}(c,d)$ for a given fixed $d$, as we increase the number of 
copies $c$. Taking the logarithm 
\bea
\ln D_{\mathrm{sym}}(c,d)
= \ln(c+d-1)! - \ln c! - \ln(d-1)!,
\label{lnD:eq}
\eea
and applying the Stirling’s formula
$\ln \eta! = \eta\ln \eta - \eta + O(\ln \eta)$,  
 to the first two factorials, we have 
\begin{align*}
\ln(c+d-1)!
&\approx (c+d-1)\ln(c+d-1) - (c+d-1) ,\\
\ln c!
& \approx c\ln c - c .
\end{align*}
Moreover, $\ln(c+d-1)=\ln c + \ln\!\big(1+\tfrac{d-1}{c}\big)
\approx \ln c + \tfrac{d-1}{c} + O(1/c^2)$, and thus 
\[
(c+d-1)\ln(c+d-1)
= (c+d-1)\ln c + (d-1) + O(1/c).
\]
Substituting back to Eq. (\ref{lnD:eq}) we obtain 
\[
\ln D_{\mathrm{sym}}(c,d)
\approx (d-1)\ln c - \ln(d-1)! + O\!\Big(\frac{1}{c}\Big).
\]
Converting to base-2 logarithms,
\[
 \log_2 D_{\mathrm{sym}}(c,d)
\approx (d-1)\log_2 c - \log_2\big((d-1)!\big) + O\!\Big(\frac{1}{c}\Big).
\]
The logarithmic dependence of
$\log_2 D_{\mathrm{sym}}(c,d)$ on the number of copies $c$
(for fixed $d$) implies that the accessible information grows only
logarithmically with $c$, rather than linearly as suggested by the
dimension of the full tensor-product Hilbert space. This observation
underlies the information-theoretic bounds discussed in Sec. \ref{sec:const3}.

For $d=2$ (qubits) the occupation numbers reduce to $(n_0,n_1)$ with $n_0+n_1=c$, 
yielding $c+1$ symmetric \emph{Dicke states}
\[
\ket{D_k^{(c)}}=
\frac{1}{\sqrt{\binom{c}{k}}}
\sum_{\pi\in S_c} \hat{P}_\pi\big(\ket{1}^{\otimes k}\otimes\ket{0}^{\otimes(c-k)}\big),
\qquad k=0,1,\dots,c,
\]
where $k$ denotes the number of excitations i.e., the number of qubits in the state $\ket{1}$.  
In angular-momentum language, the Dicke states form the spin-$j=c/2$ irreducible representation 
of $\mathrm{SU}(2)$, with basis states $\ket{j,m}$ where $m=-j,-j+1,\dots,j$, so that 
$\dim\{ \mathrm{Sym}^c(\mathbb{C}^2)\} =2j+1=c+1$, in perfect agreement with the combinatorial result.

\section{Average state for $c$ identical copies}
\label{appendix4}

For a hidden label $x$ chosen uniformly at random from ${\mathbb X}=\{0,\ldots 2^n-1\}$, the challenger prepares $c$ \emph{identical} copies
\(
\rho_x^{\otimes c} \,=\, \big(\ket{\psi_x}\!\bra{\psi_x}\big)^{\otimes c},
\)
where  $\ket{\psi_x} = (\ket{0} + e^{i\phi_x}\ket{1})/\sqrt{2}$ with $\phi_x=2\pi x/2^n$.
Because all copies are identical, $\rho_x^{\otimes c}$ is supported entirely on the
\emph{symmetric subspace} $\mathrm{Sym}^c(\mathbb C^2)$ of dimension $c+1$.
Let $\left \{\,\left | D_k^{(c)}\right \rangle \,\right \}_{k=0}^c$ be the Dicke basis (with $k$ excitations) for this subspace.
One may express the qubit states in the Dicke basis obtaining 
\[
\ket{\psi_x}^{\otimes c}
\;=\;
\sum_{k=0}^c \sqrt{\binom{c}{k}}\;\frac{e^{ik\phi_x}}{2^{c/2}}\,\ket{D_k^{(c)}},
\]
and hence
\[
\rho_x^{\otimes c}
\;=\;
\sum_{k,k'=0}^c
\frac{\sqrt{\binom{c}{k}\binom{c}{k'}}}{2^{c}}\,e^{i(k-k')\phi_x}\;
\ket{D_k^{(c)}}\!\bra{D_{k'}^{(c)}}.
\]

From the adversary's perspective, the relevant object is the ensemble average
\[
\bar{\rho}^{(c)} \;=\; \frac{1}{|\mathbb X|}\sum_{x\in\mathbb X} \rho_x^{\otimes c}.
\]
For $0\le k,k'\le c$ and $|\mathbb X|=2^n>c$, we have  the discrete orthogonality
$\frac{1}{|\mathbb X|}\sum_{x} e^{i (k-k')\phi_x} = \delta_{k,k'}$, and thus 
all off-diagonal terms vanish yielding  the diagonal Dicke decomposition
\begin{equation}
\bar{\rho}^{(c)} \;=\;
2^{-c}\sum_{k=0}^c \binom{c}{k}\ket{D_k^{(c)}}\!\bra{D_k^{(c)}}.
\label{eq:bar-copies}
\end{equation}
The diagonal form (\ref{eq:bar-copies}) implies that the rank of
$\bar{\rho}^{(c)}$ is at most $c+1$, which is the key observation
underlying the logarithmic dependence of the accessible information
on the number of copies discussed in Sec. \ref{sec:const3}.

Equation (\ref{eq:bar-copies}) can be interpreted as a discrete version of
$U(1)$ twirling around the $z$ axis. Averaging over the phases
$\phi_x$ removes coherences between Dicke states with different
excitation numbers $k$, while preserving the corresponding
populations. As a result, the averaged state is diagonal in the
Dicke basis with the binomial spectrum shown in
Eq.~(\ref{eq:bar-copies}).
By contrast, averaging over all single-qubit unitaries according
to the Haar measure ($U(2)$ twirling) yields the maximally mixed
state on the symmetric subspace,
\[
\frac{\hat{\Pi}_{\mathrm{sym}}}{c+1},
\]
where $\hat{\Pi}_{\mathrm{sym}}$ denotes the projector onto the symmetric subspace 
$\mathrm{Sym}^c(\mathbb{C}^2)$.  
In the Dicke basis this state
corresponds to a uniform distribution over all Dicke states,
\[
\frac{\hat{\Pi}_{\mathrm{sym}}}{c+1}
=
\frac{1}{c+1}
\sum_{k=0}^{c}
|D_k^{(c)}\rangle\!\langle D_k^{(c)}|.
\]
Finally, the Dicke weights $2^{-c}\binom{c}{k}$ concentrate around $k\approx c/2$ with standard deviation $\sqrt{c}/2$.

It is only in the case of $c = 1$, that one obtains the maximally mixed state 
$\bar{\rho}^{(1)}=\hat{\mathbb I}_2/2$, so the ensemble is
information-theoretically indistinguishable from uniform noise.
For $c\ge2$, the challenger still prepares $c$ copies of a \emph{pure} symmetric state $\rho_x^{\otimes c}$,
but the averaged state seen by an  adversary is the binomial mixture~\eqref{eq:bar-copies}.
Thus, $\bar{\rho}^{(c)}$ is not indistinguishable from the fully maximally mixed state $\hat{\mathbb I}_2^{\otimes c}/2^c$. 
The distinguishing measurement consists of the two-outcome POVM 
$\{\hat{\Pi}_{\mathrm{sym}},\,\hat{\mathbb I}-\hat{\Pi}_{\mathrm{sym}}\}$. This test accepts $\bar{\rho}^{(c)} $ with probability 1, while it accepts $\hat{\mathbb I}^{\otimes c}/2^c$ with probability $(c+1)/2^c$, yielding advantage $1-(c+1)/2^c$
%A simple test that measures whether the state lies in the symmetric subspace distinguishes them with advantage $1 - (c+1)/2^c$.
The appropriate reference for indistinguishability at $c\ge2$ is the \emph{symmetric} binomial mixture \eqref{eq:bar-copies}.

\section{How close is the discrete phase average to the diagonal state?}
\label{appendix5}

Let  \[\bar\rho^{(1)}=\frac{1}{N}\sum_{x=0}^{N-1}\ket{\sqrt{\mu} e^{{\rm i}2\pi x/N}}\!\bra{\sqrt{\mu} e^{{\rm i}2\pi x/N}}\] 
be the discrete phase average, and let 
\[ \bar\sigma^{(1)}=\frac{1}{2\pi}\int_0^{2\pi} d\phi\,\ket{\sqrt{\mu} e^{{\rm i}\phi}}\!\bra{\sqrt{\mu} e^{{\rm i}\phi}}
=\sum_{n\ge0} {\mathscr P}(\mu,n) \ket{n}\!\bra{n}\] 
be the fully phase-randomized (diagonal) state, with the Poisson distribution  
${\mathscr P}(\mu,n) := e^{-{\mu}} \mu^n/n!$.

Using the expansion (\ref{alpha_fock:exp}), and the expression for the sum of the finite geometric series one can show that 
\bea
\bar{\rho}^{(1)} &=& e^{-\mu} \sum_{n=0}^{\infty}\sum_{n^\prime=0}^{\infty} 
\frac{\mu^{(n+n^\prime)/2}}{\sqrt{n!n^\prime!}}\ket{n}\bra{n^\prime} 
\delta(|n-n^\prime|=q N)
\nonumber\\ 
&=&  \sum_{n=0}^{\infty}\sum_{n^\prime=0}^{\infty} 
\sqrt{{\mathscr P}(\mu,n){\mathscr P}(\mu,n^\prime)}
\ket{n}\bra{n^\prime} 
\delta(|n-n^\prime|=q N), 
\eea
for $q\in \mathbb{N}_0$, where $\delta(\cdot)$ has non-zero contributions 
for $|n-n^\prime|=0,N,2N, \ldots$ only.

The Poisson distribution is discrete and for $\mu<1$ non-negligible 
probabilities mainly exist for values of $n$ in the interval $[0, n_{\rm max}]$, with $n_{\rm max} := \lceil \mu+3\sqrt{\mu} \rceil$. 
If $N$ is much larger than the photon-number values carrying appreciable Poisson weight, 
the only relevant nonzero contributions occur for $n=n'$, and the discrete phase average 
is practically indistinguishable from the fully phase-randomized diagonal state 
\bea
\bar{\rho}^{(1)} \simeq e^{-\mu}  \sum_{n=0}^{\infty} 
\frac{\mu^n}{n!}\ket{n}\bra{n} :=\bar{\sigma}^{(1)} . 
\label{rho_diag:eq}
\eea  
The trace-distance between $\bar{\rho}^{(1)} $ and $\bar{\sigma}^{(1)} $ is essentially determined by the probability for having $|n-n^\prime| >0 $.  
For each $q\geq 1$, the density matrix $w = \bar{\rho}^{(1)}  - \bar{\sigma}^{(1)} $ contains only off-diagonal elements between states 
$\ket{n}$ and $\ket{n\pm qN}$.  In particular, we have $w = \sum_{q = 1}^\infty w^{(q)}$, where 
\[w^{(q)} = \sum_{n=0}^{\infty} 
\sqrt{{\mathscr P}(\mu,n){\mathscr P}(\mu,n+qN)} \left( \ket{n}\bra{n+q N} +  \ket{n+q N}\bra{n} \right ),\]
is Hermitian and acts only on the subspace of Fock states that differ by $qN$. 
Using triangle inequality, we have for the trace distance 
\bea
D(\bar{\rho}^{(1)} , \bar{\sigma}^{(1)} ) &=&  \frac{1}2 || w  ||_1 \leq  \frac{1}2 \sum_{q = 1}^\infty || w^{(q)}  ||_1 
\leq  \sum_{q = 1}^\infty \sum_{n=0}^\infty \sqrt{{\mathscr P}(\mu,n){\mathscr P}(\mu,n+qN)}  \nonumber\\
&&\leq e^{-\mu} \sum_{q = 1}^\infty \left( \frac{e \mu}{qN} \right )^{qN}\leq e^{-\mu} \sum_{q = 1}^\infty \left( \frac{e \mu}{N} \right )^{qN},
\eea 
 where the Chernoff-Hoeffding bound has been used for ${\mathscr P}(\mu,n+qN)$, while we have focused on cases $N\geq 4$ and $\mu<1$. 
 Using the summation of the geometric series on the right-hand side we obtain
\bea
D(\bar{\rho}^{(1)} , \bar{\sigma}^{(1)} )  \leq 
e^{-\mu}\frac{ \left( \frac{e \mu}{N} \right )^N}{1- \left( \frac{e \mu}{N} \right )^N }, 
 \eea
which drops faster than exponentially in $N$, for $N\gg 1$.

\section{Operational interpretation of the trace distance}
\label{appendix7}
The following result, often referred to as Helstrom's theorem,
provides the operational interpretation of the trace distance.
%It states that the trace distance between two quantum states equals
%the maximum distinguishing advantage achievable by any binary
%measurement.

For any density operators $\rho$, and $\sigma$ we have 
\[
\max_{\{\hat{M}_0,\hat{M}_1\}} \Big|\mathrm{Tr}(\hat{M}_1 \rho)-\mathrm{Tr}(\hat{M}_1 \sigma)\Big|
= \frac{1}{2}\,\|\rho-\sigma\|_1,
\]
where the maximum is over all binary POVMs $\{\hat{M}_0,\hat{M}_1\}$ with $\hat{M}_0+\hat{M}_1=\hat{\mathbb I}$. 
Operationally, the left-hand side equals the maximum difference
between the probabilities of obtaining a chosen measurement outcome
when the system is prepared in state $\rho$ or in state $\sigma$.
This quantity is commonly referred to as the distinguishing advantage.

\begin{proof}
Let $\Delta := \rho - \sigma$, which is Hermitian and satisfies $\mathrm{Tr}(\Delta)=0$. Consider now the spectral decomposition of $\Delta$
\bea
\Delta = \sum_i \lambda_i \ket{i}\bra{i}
\eea
and define
\bea
\Delta^+ = \sum_{\lambda_i>0} \lambda_i \ket{i}\bra{i},\quad 
\Delta^- = 
\sum_{\lambda_i<0} (-\lambda_i) \ket{i}\bra{i}. 
\eea
Then 
\[
\Delta = \Delta^{+} - \Delta^{-},
\]
 and $\Delta^{+}\Delta^{-}=0$ (orthogonal supports). 
The eigenvectors corresponding to $\lambda_i=0$ belong to neither
$\Delta^+$ nor $\Delta^-$, but their contribution vanishes.

By definition of the trace norm for Hermitian operators,
\[
\|\Delta\|_1 = \mathrm{Tr}(\Delta^{+}) + \mathrm{Tr}(\Delta^{-}).
\]
Since $\mathrm{Tr}(\Delta)=0$, it follows that
\begin{equation}\label{eq:half-split}
\mathrm{Tr}(\Delta^{+}) = \mathrm{Tr}(\Delta^{-}) = \tfrac12 \|\Delta\|_1.
\end{equation}

Let $\{\hat{M}_0,\hat{M}_1\}$ be any binary POVM. Then 
\[
\mathrm{Tr}(\hat{M}_1\Delta)
= \mathrm{Tr}(\hat{M}_1\Delta^{+}) - \mathrm{Tr}(\hat{M}_1\Delta^{-})
\le \mathrm{Tr}(\Delta^{+})
= \tfrac12 \|\Delta\|_1,
\]
because $0\le \hat{M}_1\le \hat{\mathbb I}$ and $\Delta^\pm \ge 0$ imply $0\le \mathrm{Tr}(\hat{M}_1\Delta^{\pm}) \le \mathrm{Tr}(\Delta^{\pm})$.
Similarly,
\[
-\mathrm{Tr}(\hat{M}_1\Delta)
= \mathrm{Tr}(\hat{M}_1\Delta^{-}) - \mathrm{Tr}(\hat{M}_1\Delta^{+})
\le \mathrm{Tr}(\Delta^{-})
= \tfrac12 \|\Delta\|_1.
\]
Hence for every binary POVM,
\[
\big|\mathrm{Tr}(\hat{M}_1\Delta)\big| \le \tfrac12 \|\Delta\|_1.
\]

We will now show that the equality can be achieved when $\hat{M}_1:=\hat{P}^{+}$, where 
$\hat{P}^{+}:=\sum_{\lambda_i>0}\ket{i}\bra{i}$ is the projector onto the support of $\Delta^{+}$.
Because $\Delta^{+}\Delta^{-}=0$, we have $\hat{P}^{+}\Delta^{-}=0$ and $\hat{P}^{+}\Delta^{+}=\Delta^{+}$.  
So, using Eq. \eqref{eq:half-split} we obtain 
\[
\mathrm{Tr}(\hat{M}_1\Delta)
= \mathrm{Tr}(\hat{P}^{+}\Delta^{+}) - \mathrm{Tr}(\hat{P}^{+}\Delta^{-})
= \mathrm{Tr}(\Delta^{+})
= \tfrac12 \|\Delta\|_1,
\]
which matches the upper bound.

We have shown that no binary POVM can achieve an advantage
larger than $\tfrac12\|\rho-\sigma\|_1$, and that this value
is achieved by the measurement
$\{\hat P^+,\,\hat{\mathbb I}-\hat P^+\}$.
Therefore
\[
\max_{\{\hat{M}_0,\hat{M}_1\}}
\big|\mathrm{Tr}(\hat{M}_1 \rho)-\mathrm{Tr}(\hat{M}_1 \sigma)\big|
= \tfrac12 \|\rho-\sigma\|_1.
\]
%%%%
% Combining the two parts,
%\[
%\max_{\{\hat{M}_0,\hat{M}_1\}} \big|\mathrm{Tr}(\hat{M}_1 \rho)-\mathrm{Tr}(\hat{M}_1 \sigma)\big|
%= \tfrac12 \|\rho-\sigma\|_1.
%\]
%%%%
\end{proof}

Therefore, the trace distance between two average states directly
quantifies the maximum advantage of any physically allowed
distinguisher, which justifies Eq.~(\ref{IT_Ind_Av:Qdef}) in the main text.

\section{Haar-random unitaries and quantum states}
\label{appendix9}

The unitary group ${\mathfrak U}(d)$ consists of all $d\times d$ complex matrices
$U$ satisfying $U^\dagger U = UU^\dagger = {\mathbb I}_d$, i.e., all linear
transformations that preserve inner products on $\mathbb{C}^d$.
Equivalently, elements of ${\mathfrak U}(d)$ map orthonormal bases to orthonormal
bases. Equipped with matrix multiplication, ${\mathfrak U}(d)$ forms a compact
Lie group whose real dimension is $d^2$. In quantum mechanics, unitary
operators describe reversible time evolution of closed systems and
constitute the mathematical model for quantum gates.

To define what it means to choose a unitary ``at random'' from ${\mathfrak U}(d)$,
one uses the notion of a probability measure. Informally, a probability
measure specifies how likely different regions of a space are when
sampling an element at random, assigning probabilities to sets of
outcomes rather than to individual points\footnote{This situation is entirely analogous to a classical random
number generator that outputs a real number uniformly from $[0,1]$.
Individual outcomes occur with probability zero, while nontrivial
probabilities are assigned only to intervals. By contrast, if one
samples uniformly from the finite set of all $n$-bit strings, the sample
space is discrete, and each individual string occurs with probability
$2^{-n}$ rather than zero.}.

The \emph{Haar measure} $\mu_{\mathrm{H}}$ on the unitary group ${\mathfrak U}(d)$
is the unique probability measure that is invariant under left and right
multiplication by fixed unitaries. Concretely, for any measurable subset
$\mathcal{A} \subseteq {\mathfrak U}(d)$\footnote{Here $\mathcal A$ denotes an arbitrary measurable subset of ${\mathfrak U}(d)$,
i.e., any collection of unitaries representing an event of interest.
For example, $\mathcal A$ could be the set of all unitaries within
distance $\varepsilon$ from the identity.}, and any unitary
$V \in {\mathfrak U}(d)$, one has
\[
\mu_{\mathrm{H}}(V \mathcal{A}) = \mu_{\mathrm{H}}(\mathcal{A})
= \mu_{\mathrm{H}}(\mathcal{A} V),
\]
where $V\mathcal{A} := \{VW : W \in \mathcal{A}\}$ and
$\mathcal{A}V := \{WV : W \in \mathcal{A}\}$.

Haar-random unitaries induce a natural notion of randomness for quantum
states. A pure state $\ket{\psi}$ in $\mathbb{C}^d$ is called
\emph{Haar-random} if it is obtained by applying a Haar-random unitary
to a fixed reference state, e.g.\ $\ket{\psi}=U\ket{0}$. This procedure
induces the uniform probability distribution on the unit sphere of
$\mathbb{C}^d$. Intuitively, Haar-random states are maximally
unstructured and uniformly distributed over Hilbert space. They provide
the canonical notion of completely random pure quantum states and
possess many typical properties that hold with overwhelming probability
in high dimensions.

In the mixed-state setting, one commonly considers the
\emph{Haar-induced ensemble}. A density matrix $\rho$ on a subsystem
$\mathcal{H}_{\rm S}$ is obtained by sampling a Haar-random pure state
$\ket{\Psi}$ on the composite Hilbert space
$\mathcal{H}_{\rm S}\otimes\mathcal{H}_{\rm E}$ and tracing out the
environment,
\[
\rho=\mathrm{Tr}_{\rm E}\!\left(\ket{\Psi}\!\bra{\Psi}\right).
\]

Haar-random and Haar-induced ensembles serve as the standard reference
ensembles for quantum randomness. Pseudorandom state generators (PRSGs)
and pseudorandom density-matrix generators (PRDMs) are defined as
efficient procedures whose outputs are computationally indistinguishable
from the corresponding Haar-random pure-state or Haar-induced mixed-state
ensembles by any QPT adversary.

A central structural property underlying many results on random quantum
states is that Haar-random pure states of sufficiently
high-dimensional bipartite systems are almost maximally entangled with
overwhelming probability~\cite{Hayden2006GenericEntanglement}.

For further discussion of Haar measure and its role in quantum
information theory, see~\cite{Mele2024HaarMeasureTutorial} and the
references therein.

\bibliographystyle{iopart-num}
\bibliography{refs_qowf}
\end{document}